\documentclass[journal]{IEEEtran}

\usepackage{cite}

\usepackage{amsmath,amsfonts,amssymb}
\usepackage[caption=false]{subfig}
\usepackage{graphicx}     
\graphicspath{{./Figures/}}
\usepackage{subfig}
\usepackage{enumitem}
\usepackage{color}
\usepackage{eurosym}
\usepackage{url}
\usepackage{textcomp}
\usepackage{multirow}
\usepackage{booktabs}
\usepackage{hhline}
\usepackage{interval}
\usepackage{epstopdf}
\usepackage{caption}
\usepackage[section]{placeins}
\usepackage{mathtools}
\usepackage{commath}
\usepackage[colorlinks]{hyperref}
\usepackage[algo2e,ruled,lined,linesnumbered]{algorithm2e}

\usepackage{lineno}
\usepackage{siunitx}
\usepackage{balance}

\begin{document}
\title{Monitoring of Railpad Long-term Condition in Turnouts Using Extreme Value Distributions}
\author{Pegah~Barkhordari,~
        Roberto~Galeazzi,~\IEEEmembership{Member,~IEEE,}and~Mogens~Blanke,~\IEEEmembership{Senior Member,~IEEE}
\thanks{Pegah~Barkhordari,~Roberto~Galeazzi,~and~Mogens~Blanke are with the Automation and Control Group, Department of Electrical Engineering, Technical University of Denmark, Kgs. Lyngby DK-2800, Denmark (e-mail: pebark@elektro.dtu.dk;
rg@elektro.dtu.dk; mb@elektro.dtu.dk).}
}

\maketitle

\IEEEtitleabstractindextext{%
\begin{abstract}
The railpad is a key element in railway infrastructures that plays an essential role in the train-track dynamics. Presence of worn or defective railpads along railway track  may lead to large wheel/rail interaction forces, and a high rate of deterioration for track components. Despite the importance of railpad, the track infrastructure managers use no inspection tool for monitoring in-service railpads over time. In this paper, a novel data-driven monitoring tool for long-term performance analysis of in-service railpads is developed based on train-induced vibration data collected by a track-side measurement system. The monitoring tool consists of a method for track resonance frequencies estimation, a temperature-frequency model for describing railpad behavior with respect to ambient temperature, and a generalized likelihood ratio test based on the generalized extreme value distribution for detecting changes in the railpad status over time. To evaluate the performance of the proposed monitoring system, the status of railpads at four different locations along a railway turnout is monitored over a period of 18 months. It is shown that the monitoring system can successfully detect changes in railpad properties over the considered period.

\end{abstract}

}

\maketitle

\IEEEdisplaynontitleabstractindextext

%
\IEEEpeerreviewmaketitle

\section{Introduction}
%
%
%
%
\IEEEPARstart{P}{erformance} of turnouts as vital elements of railway infrastructure significantly influences the safety and capacity of railway networks. Reliability and availability of a railway turnout strongly depend on the long-term condition of its components. Among the different components, ballast and railpad are key elements which play an essential role in dynamic behavior of railway track along the turnout. \\
The railpad is an elastic layer placed between the rail and sleepers which attenuates wheel/rail interaction forces. The stiffness of the railpad is a significant parameter affecting the track deterioration rate and the train-induced noise and vibration level \cite{thompson2008railway}. Railway tracks with stiff railpads typically experience large wheel/track interaction forces and a relatively high level of rolling noise \cite{thompson1995track}. Soft railpads, on the other hand, result in a lower interaction force and rolling noise. However, they can accelerate the deterioration of track components \cite{oregui2016obtaining}. It is therefore important to not only determine an optimal range for the stiffness of railpads at the design stage of railway tracks, but also monitor the long-term variations of the stiffness for railpads in service. The latter needs a comprehensive analysis of field data through a method which takes into account different influencing parameters. There are several factors impacting the railpad stiffness, such as preload, temperature and aging \cite{oregui2016obtaining}. Effect of preload on the dynamic characteristics of railpads has been investigated in a number of studies by performing laboratory tests \cite{thompson1998developments, kaewunruen2008alternative, oregui2016obtaining, maes2006measurements}. These studies confirm the dependency of railpad stiffness on preload, and indicate that the stiffness increases with increasing the applied preload. In some research studies, laboratory measurements have been made in order to examine the behavior of railpads at different temperatures. Dependence of railpad properties on temperature was studied in \cite{broadbent2009evaluation} and \cite{squicciarini2016effect} by measuring variations in the stiffness of two railpad materials over a temperature range of \ang{-20}C to \ang{40}C. A clear increase in the stiffness with decreasing temperature was observed for both railpads. A combination of dynamic mechanical analysis (DMA) and the time-temperature superposition principle was used in \cite{oregui2016obtaining} to investigate the dynamic properties of three different types of railpads under temperature-controlled conditions. Based on the obtained results, it was concluded that the stiffness of railpads and, therefore, dynamic behavior of railway track vary between hot and cold days. A nonlinear behavior was observed in \cite{wei2017investigation} for three different railpad materials tested at temperatures between \ang{-40}C to \ang{70}C. The stiffness was shown to be more significantly sensitive to temperature variation below \ang{10}C. Unlike the aforementioned factors, aging causes permanent changes in railpad stiffness. The effect of aging has been investigated by researchers through comparing the dynamic properties of new and worn railpads. An instrumented impact hammer and a single degree of freedom model were used in \cite{remennikov2006deterioration} in order to examine the frequency response function of railpads at different ages. Complex stiffness of new and worn FC9 railpads were obtained in \cite{oregui2016obtaining} by performing laboratory tests. The complex stiffness of the worn railpad was found to be significantly smaller than the stiffness of the new one. A similar study was carried out in \cite{oregui2017sensitivity} and dynamic response of railway tracks with new and worn railpads obtained from a receptance test were compared. Based on the obtained results, it was concluded that the antiphase movement of rail and sleeper in the track with worn pads occurs at lower frequencies.

\subsection{Main contribution} 
This paper proposes a novel data-driven monitoring tool based on statistical analysis of railpad characteristics over time. As discussed earlier, changes in dynamic properties of railpad can be caused by different factors. For detecting deficiencies in the performance of in-service railpads, it is therefore of high importance that the monitoring tool can distinguish permanent variations of the stiffness associated with deterioration of railpad from variations caused by the other factors. This paper is intended to provide a tool with such a capability through minimizing the preload and temperature effects in the long-term performance analysis of railpads. In order to develop the monitoring tool, acceleration response of the track measured at different locations along a railway turnout by using a track-side measurement system is exploited. The data collected over a period of 18 months from September 2017 to February 2019 is considered for the analysis. The method presented in~\cite{barkhordari2018safe} is employed to extract the second track resonance frequency representing the dynamic behavior of the railpad layer. Effect of the preload is minimized by considering a pool of data corresponding to the passage of identical trains with a low level of variation in the axle load. A model describing temperature dependence of the railpad properties is then obtained by utilizing temperature data and the estimated resonance frequencies. This model is used for the purpose of minimizing the effect of temperature variation on the estimated track resonance frequency. A change detection algorithm is then designed to monitor the quality of the railpad layer. The novel data-driven tool developed in this study is robust to uncertainties present in the train-induced vibration data, and ambient temperature variations.

The paper is organized as follows: Section \ref{sec:setup_railpad} describes the track-side measurement system used to collect the data. Section \ref{sec:method_railpad} presents an overview of the proposed monitoring method. Section \ref{sec:temp} discusses effects of preload and temperature variations on the estimated frequencies. Section \ref{sec:monitor_railpad} provides a detailed description of the monitoring system design. Section \ref{sec:Impelement_railpad} examines the performance of the monitoring tool, while Section \ref{sec:discussion} discusses the detection problem at hand and the adoption of the monitoring system by railway infrastructure managers. Last, Section \ref{sec:conclusion_railpad} provides concluding remarks.

\section{Track-side measurement system} \label{sec:setup_railpad}
\begin{figure}[tbp]
	\centering
	\includegraphics[width=\linewidth]{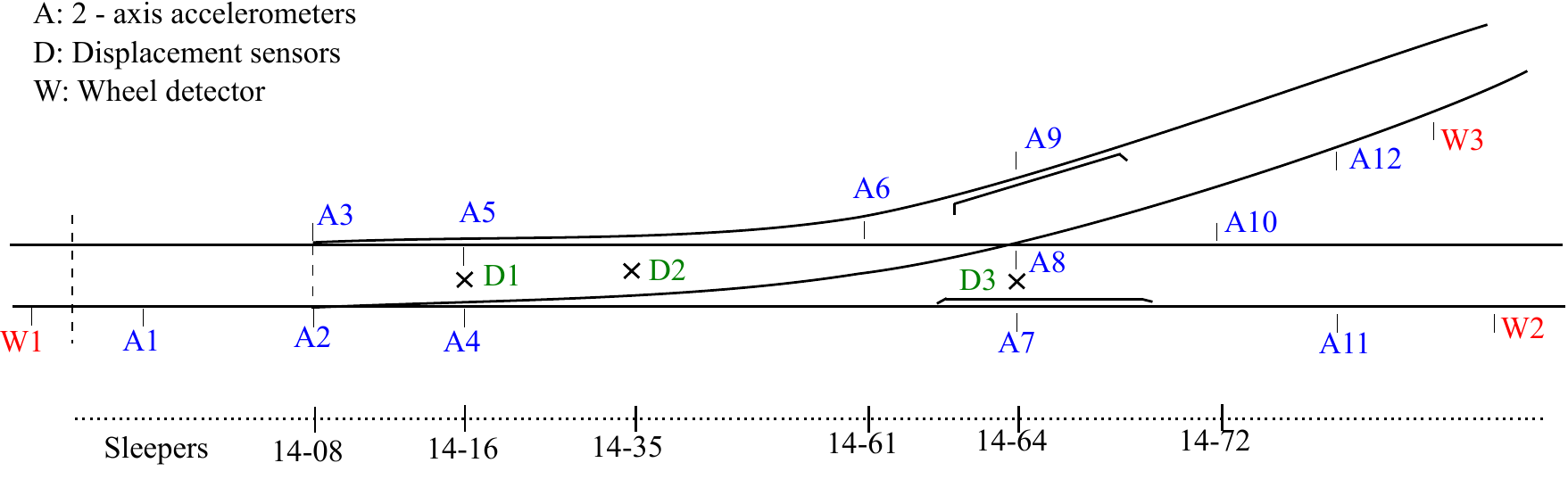}
	\caption{Layout of the sensors location along the turnout at Tommerup station (Fyn - Denmark). The data utilized in this work refer to vertical accelerations measured by accelerometers A2,A4, A7 and A11.}
	\label{fig:TommerupSensorPlacement_railpad}
\end{figure}

\begin{figure}[tbp]
	\subfloat[Accelerometer A5 magnetically connected to the rail web]{%
		\includegraphics[clip,width=0.45\columnwidth]{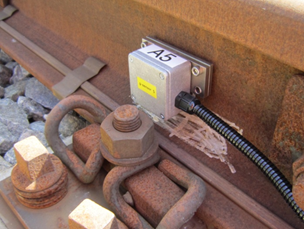}%
	}
	\hspace{0.4cm}
	\subfloat[Wheel detector sensor mounted in proximity of the rail web]{%
		\includegraphics[clip,width=0.45\columnwidth]{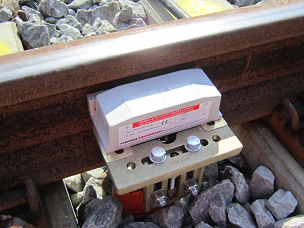}%
	} \\
	\caption{Sample pictures of the instrumentation installed along the turnout at Tommerup station (Fyn - Denmark).}
	\label{fig:TommerupSC_railpad}
\end{figure}

The intelligent data-driven tool for monitoring the railpad status over time exploits the data collected by a track side measurement system located at a turnout of the Danish railway network. As shown in Fig.~\ref{fig:TommerupSensorPlacement_railpad}, the turnout is instrumented with 12 2-axis accelerometers (measurement range: $\pm500$g), 3 displacement sensors (measurement range: $\pm20$mm) and 3 wheel detectors. All sensors are connected to a data acquisition system, where the data is band-pass filtered, amplified and temporarily stored. The wheel detectors are employed to activate the measurement system for automatic data recording during the passage of each train through the turnout. Vertical and lateral accelerations of the rail are measured using the 2-axis accelerometers magnetically installed on the rail web. Displacement sensors are mounted on the sleepers to measure their vertical displacement. The sampling frequency is set to $F_s = 20\,\mathrm{kHz}$ in the data acquisition system. Pictures of a selected accelerometer and a wheel detector are shown in Fig.~\ref{fig:TommerupSC_railpad}. 
   
\begin{figure}[tbp]
	\begin{center}
		\includegraphics[width=\linewidth]{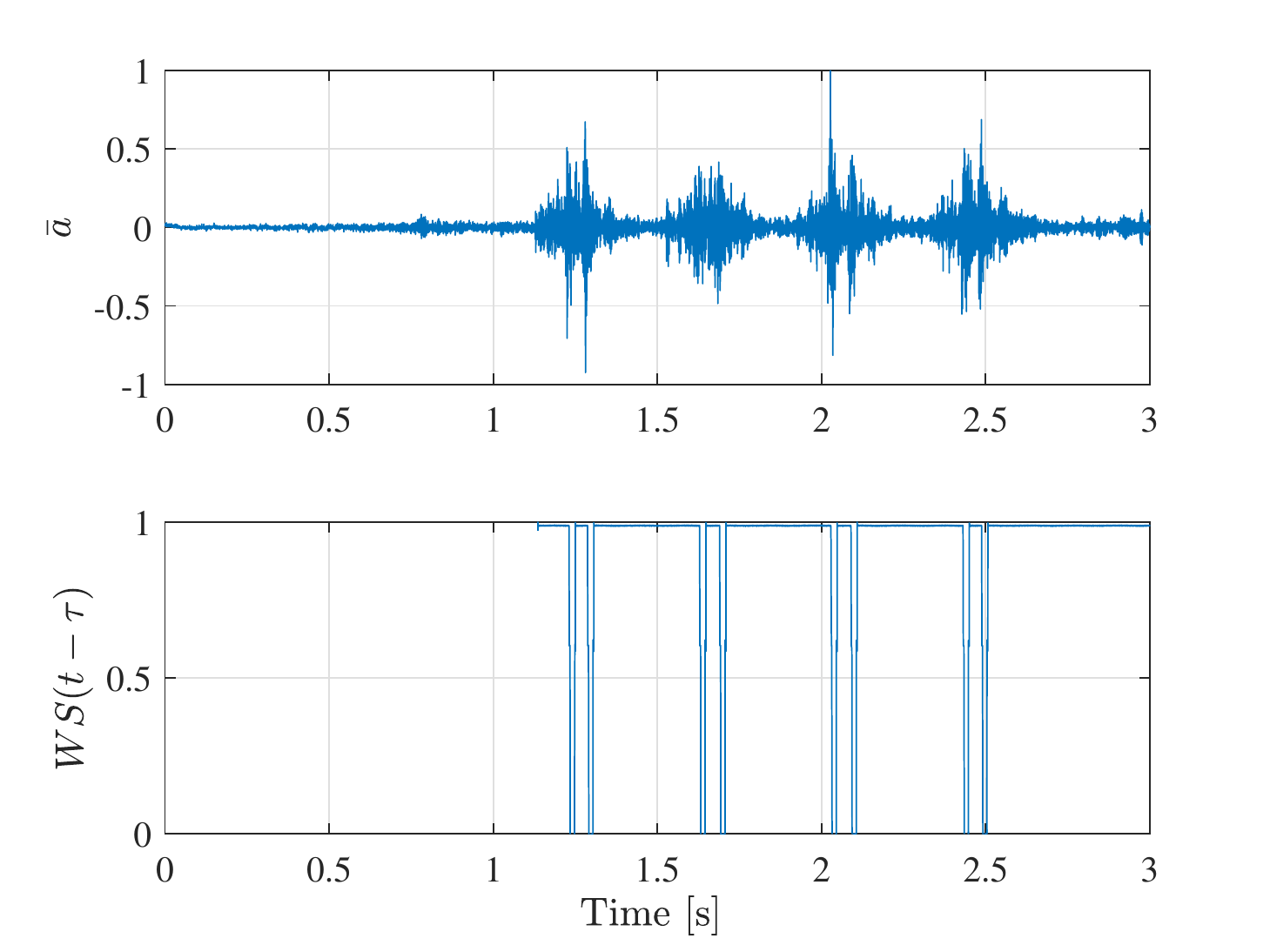}    
		\caption{The track vertical acceleration measured at location A7 in response to a IC3 passenger train and the time-shifted wheel detector signal.} 
		\label{fig:ic3A7}
	\end{center}
\end{figure}

As an example, Fig.~\ref{fig:ic3A7} shows the normalized vertical acceleration and the time-shifted output of the wheel detector recorded at location A7 during the passage of an IC3 train. The normalized acceleration is obtained as $\bar{a}=a/\max(|a|)$. Measured data presented in this paper is anonymized to comply with the policy of the Danish railway infrastructure manager.
In order to find the time instants when the train wheels pass through a measurement location along the turnout (A2, A4, A7 or A11), the output of the wheel detector, $WS(t)$, is shifted in time (i.e., synchronized with the measured acceleration $\bar{a}(t)$). The time shift $\tau$ is calculated by utilizing the train speed and the known distances between the wheel detector and the accelerometers.

An axle load checkpoint (ALC) system is installed in a nearby station located about 7 km from Tommerup station. This system provides information about the axle loads of the passing trains, which is utilized to choose trains with relatively low axle loads and minimum axle load variation, appropriate for the analysis carried out in this paper. This choice is further discussed in Section~\ref{sec:temp}.

\section{Method overview} \label{sec:method_railpad}
\begin{figure}[tbp]
    \centering
    \includegraphics[width=\linewidth]{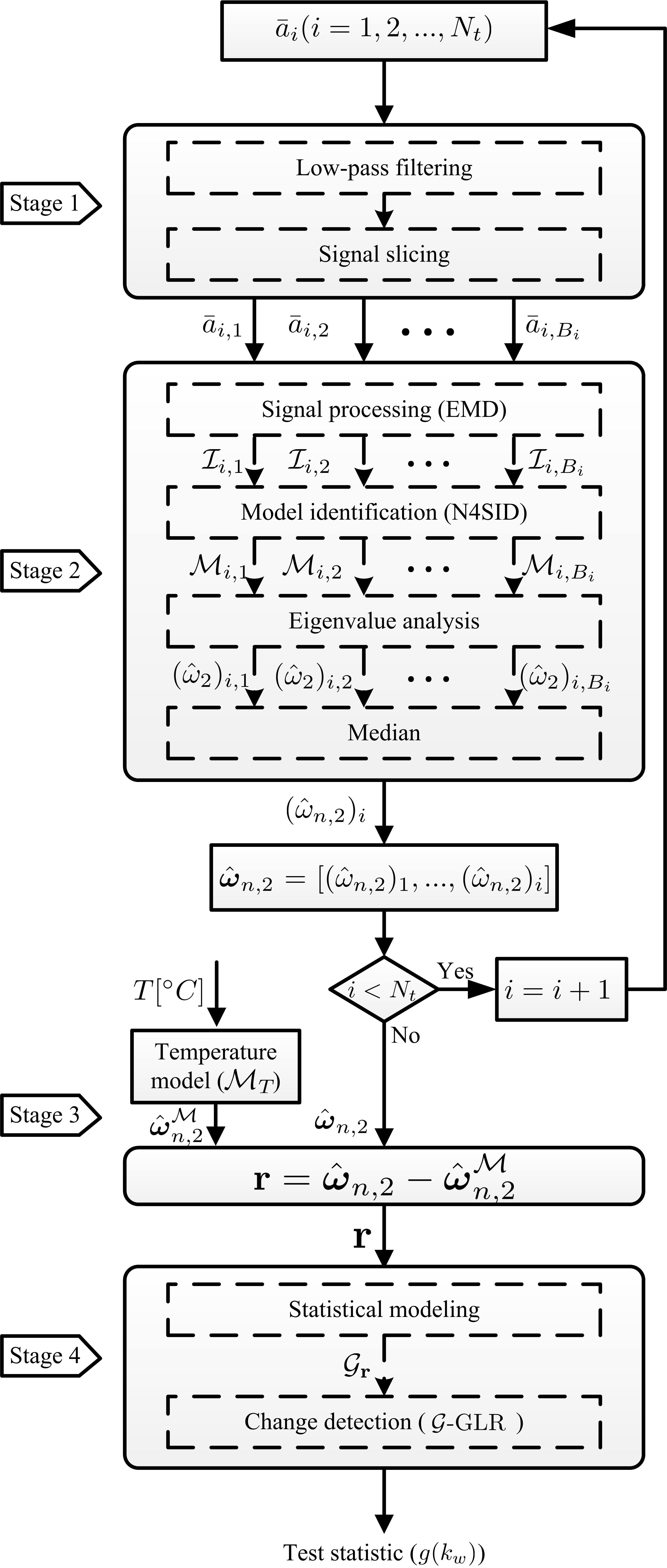}
    \caption{Proposed method for monitoring the railpad status}
    \label{fig:railpadmonitoring}
\end{figure}

The monitoring system consists of four stages: (i) the pre-processing of the train-induced vibration data including low-pass filtering and signal slicing; (ii) the extraction of the track vibration characteristics (i.e., resonance frequency and damping) representing the railpad behavior, by employing a combination of an advanced signal processing technique and a subspace identification method; (iii) the generation of a residual signal by temperature compensation of the estimated resonance frequency; (iv) the monitoring of the railpad status by means of a statistical change detection algorithm. A schematic of the monitoring strategy is shown in Fig.~\ref{fig:railpadmonitoring}.

\begin{figure}[tbp]
	\begin{center}
		\includegraphics[width=\linewidth]{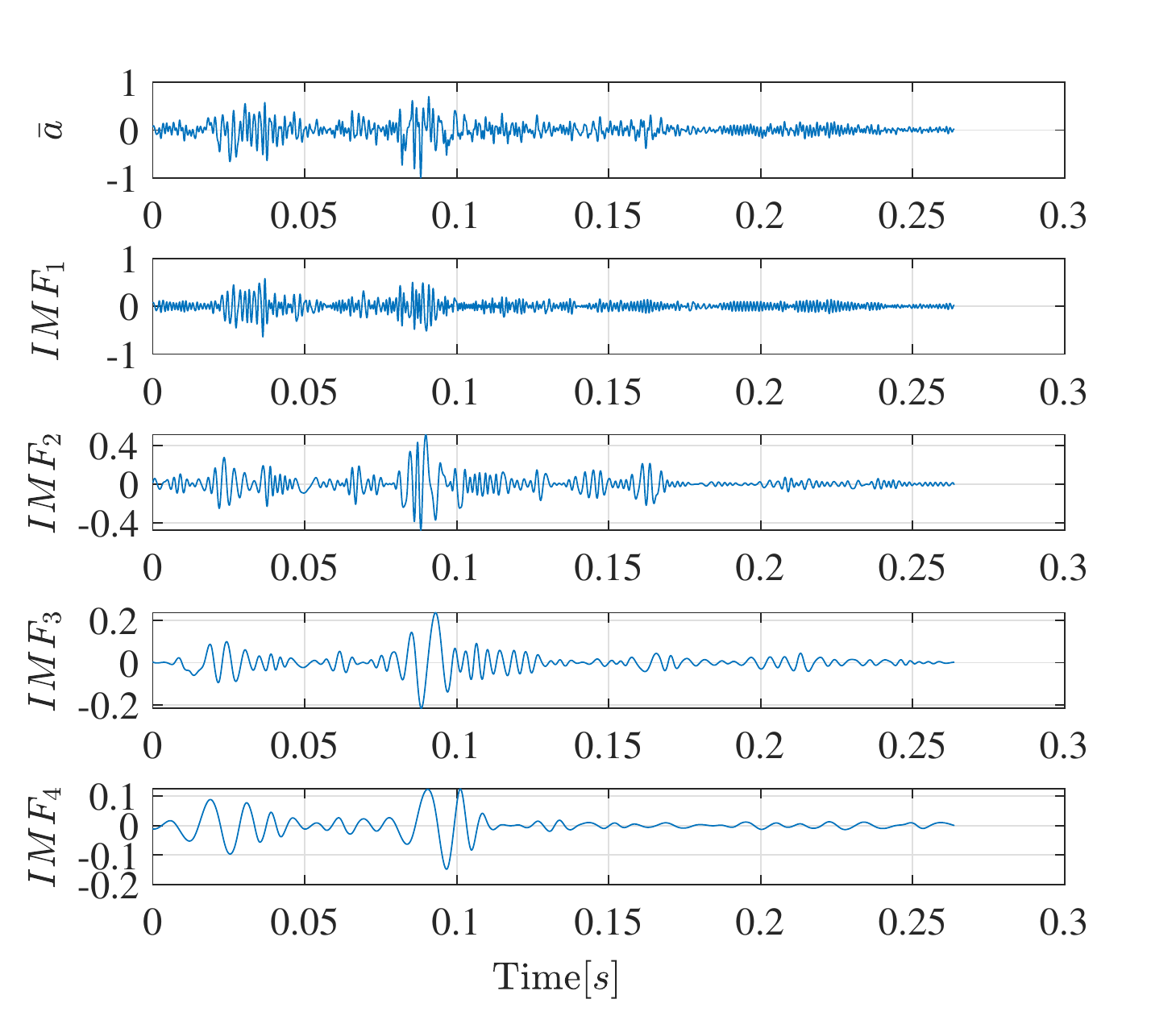}    
		\caption{The sliced track acceleration in response to a bogie passage   and the extracted IMFs.} 
		\label{fig:IMFA7_railpad}
	\end{center}
\end{figure}

The acceleration signals $\bar{a}_i$ collected by the track-side measurement system during the passage of $N_t$ IC3 trains are fed to the algorithm. In the pre-processing block, the signals are first low-pass filtered using a 3rd order Butterworth filter with a cut-off frequency of $1000 \mathrm{Hz}$. This cut-off frequency is chosen based on the frequency interval in which the dynamic behavior of the turnout in relation to the ballast and railpad layers is expected~\cite{barkhordari2017low}. The filtered signals are then sliced at every two adjacent wheels corresponding to the passage of each bogie. The signal segments obtained in the pre-processing stage (i.e., $\bar{a}_{i,1}, \bar{a}_{i,2}, ..., \bar{a}_{i,B_i}$, where $B_i$ is the number of bogies of the $i$-th train) are used as inputs to the next stage where the Empirical Mode Decomposition (EMD) algorithm is adopted for signal processing. The intrinsic oscillatory modes ($\mathcal{I}= $ (IMF$_1$, IMF$_2$, ...)) of each signal segment are extracted using the EMD algorithm. Figure~\ref{fig:IMFA7_railpad} provides an example of the obtained signal segment and the corresponding extracted IMFs at location A7.

The frequency content of the second IMF (IMF$_2$) lies within the interval in which the second track resonance frequency, corresponding to the antiphase vibrations of the rail and sleeper on the flexibility of the railpad layer, is expected \cite{barkhordari2018safe}. The subspace state space identification algorithm (N4SID) is adopted to identify a model for IMF$_2$. Since each IMF contains one dominant oscillatory mode of the acceleration signal, a model of order 2 is adequate to represent its dominant behavior. For the second IMF extracted from the $s$-th segment of the $i$-th acceleration signal $\bar{a}_{i,s}$, the canonical realization of the discrete-time state space model is given as follow,
\begin{equation}
 \mathcal{M}_{i,s}:\left\lbrace
\begin{aligned}
\hat{\mathbf{x}}_{i,s}(k+1)= \hat{\textbf{A}}_{i,s}\hat{\mathbf{x}}_{i,s}(k) \\
\hat{\mathbf{y}}_{i,s}(k)=\hat{\textbf{C}}_{i,s} \hat{\mathbf{x}}_{i,s}(k)
\end{aligned}\right. .
\end{equation}
The resonance frequency corresponding to the model identified for the IMF$_2$ can be computed by means of eigenvalue analysis,
\begin{equation}
(\hat{\omega}_{2})_{i,s} = \frac{\mid{\ln({\lambda}_1(\mathbf{\hat{A}}_{i,s}))\mid}}{2{\pi}T_s}, \quad  \label{eq:freq_damp_est}
\end{equation}
where $\lambda_1(\hat{\mathbf{A}}_{i,s})$ is the eigenvalue of the complex pair associated with the matrix~$\hat{\mathbf{A}}_{i,s}$ and $T_s$ is the sampling time.
To achieve a robust estimation procedure, a single estimation of the second track resonance frequency $(\hat{\omega}_{n,2})_i$ is computed for each train passage by taking the median over all estimated track resonance frequencies obtained from the identified models. The first and the second stages of the proposed method are thoroughly explained in~\cite{barkhordari2018safe}.

In the third stage of the method, a temperature-frequency model $\mathcal{M}_T$ describing the relation between the estimated second track resonance frequencies and ambient temperature is obtained. Hourly temperature data is used as input to this model, and a residual sequence $\textbf{r}$ in which the temperature effect is minimized is calculated based on the output of the model ($\hat{\boldsymbol{\omega}}^{\mathcal{M}}_{n,2}$) and the vector of the estimated frequencies ($\hat{\boldsymbol{\omega}}_{n,2}$). This stage is discussed in details in Section~\ref{sec:temp}. In the final stage, a statistical distribution is fitted to the obtained residual sequence. The generalized likelihood ratio test based on the generalized extreme value distribution ($\mathcal{G}$-GLRT) is then exploited for detecting changes in the distribution of the residual sequence over time.

\section{Preload and temperature effects}\label{sec:temp}
As discussed earlier, there is a number of factors that affects the properties of the railpad and, consequently, the second track resonance frequency. To properly monitor the railpad quality over time, it is necessary to distinguish between the changes in the railpad properties associated with the preload and temperature effects and the changes linked to the aging/deterioration effect. The effects of the preload and temperature on the estimated frequencies are first analyzed; then the strategies adopted for minimizing these effects are presented.

\subsection{Preload effect}
\begin{figure}[tbp]
	\begin{center}
		\includegraphics[width=\linewidth]{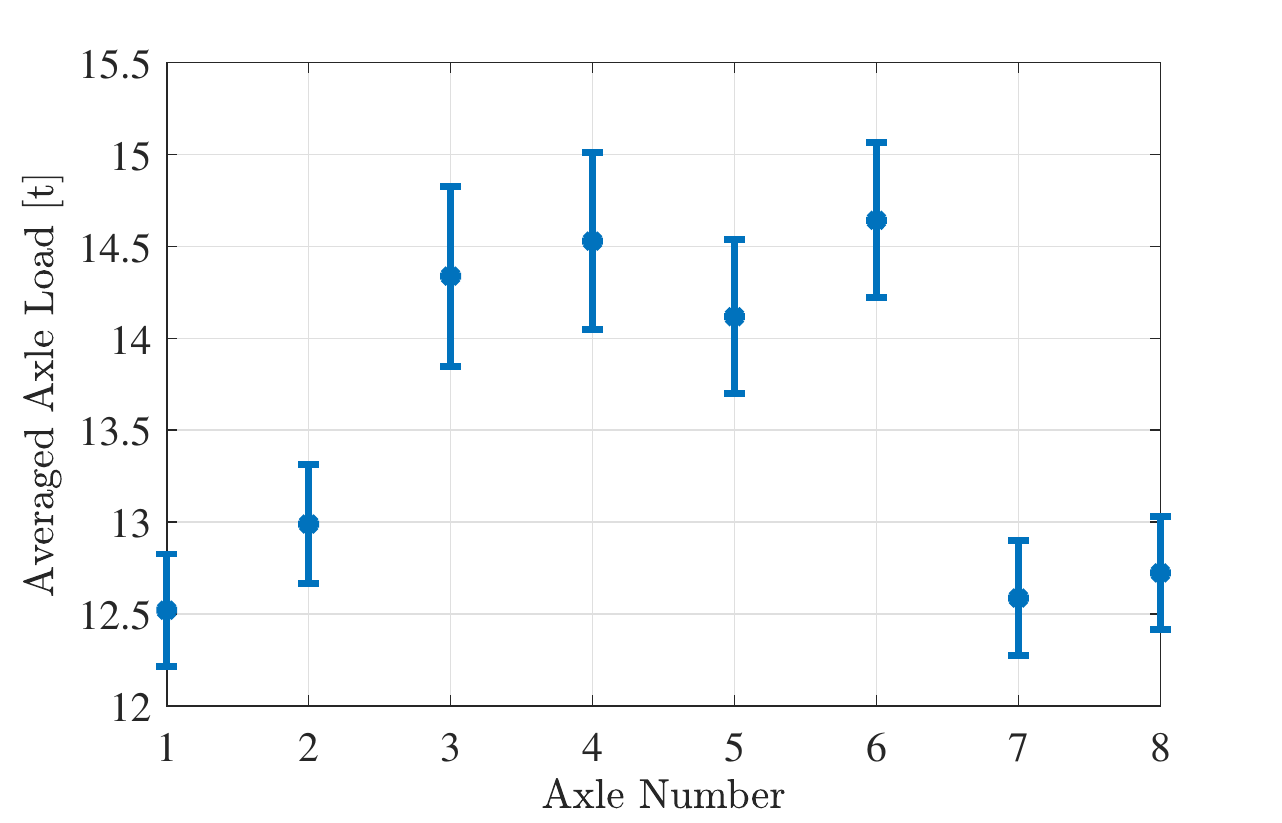}    
	\caption{Mean-value of the axle loads and their 1 $\sigma$ standard deviation over a period of one month.} 
		\label{fig:IC3load_railpad}
	\end{center}
\end{figure}

According to the analysis performed in \cite{Wu1999effects}, wheel preload applied to a railpad significantly influences its stiffness. The stiffness of the railpad has been shown to increase with increasing the wheel preload. 
Since different types of trains with a variety of wheel loads are passing through the turnout, a considerable variation is observed in the estimated values of the second track resonance. In order to attenuate the effect of wheel load variations, only the track acceleration signals induced by a batch of identical IC3 trains with minimum wheel load variation are considered in the analysis. Figure~\ref{fig:IC3load_railpad} shows the averaged axle loads and the 1$\sigma$ standard deviation for IC3 trains which have passed through the turnout over a period of one month (September 2017).

\begin{figure}[tbp] 
	\centering
	\subfloat[ A2 ]{\includegraphics[width=0.49\columnwidth]{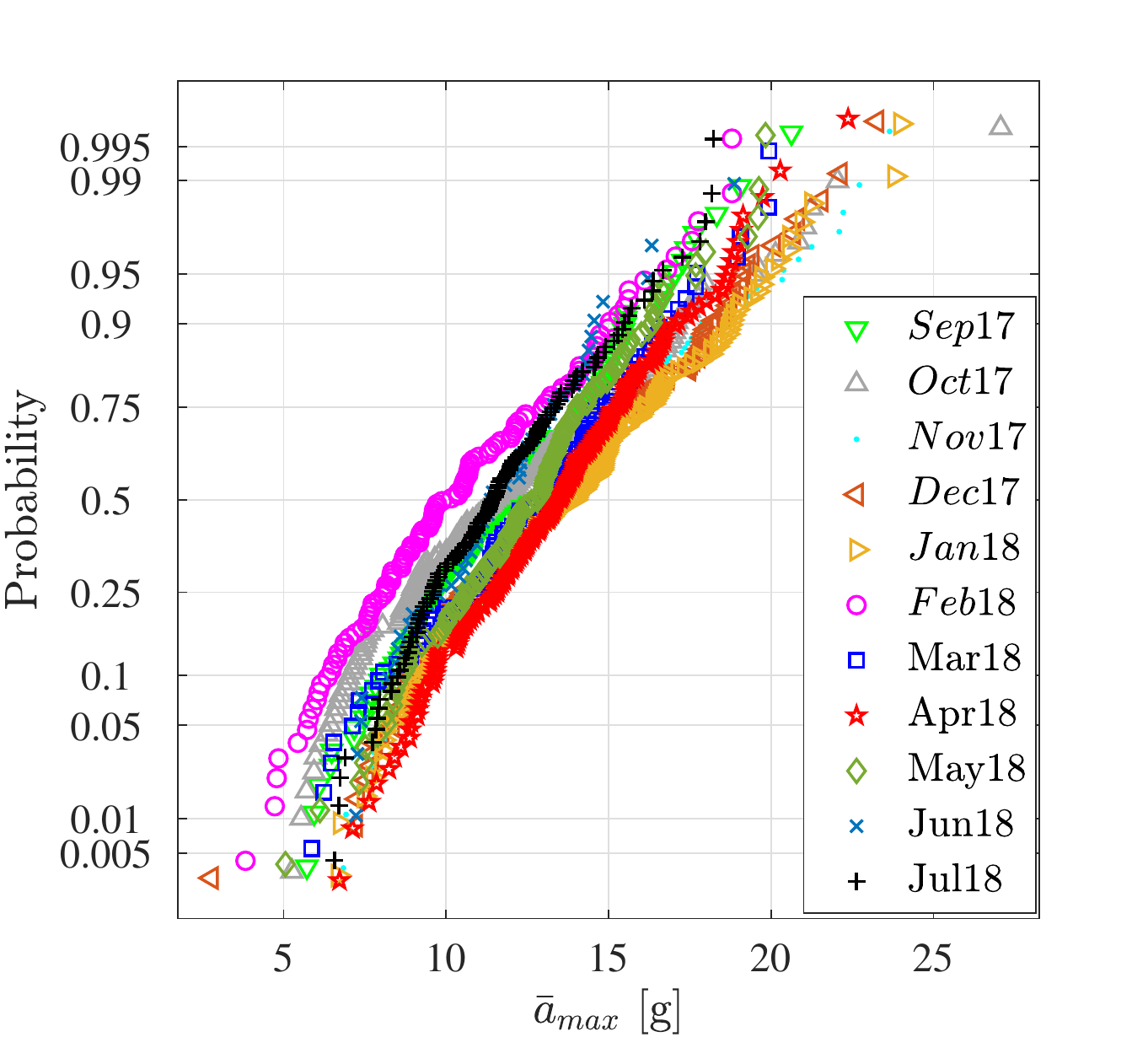}%
		\label{}}
	\hspace{0.05cm}	
	\subfloat[A4]{\includegraphics[width=0.49\columnwidth]{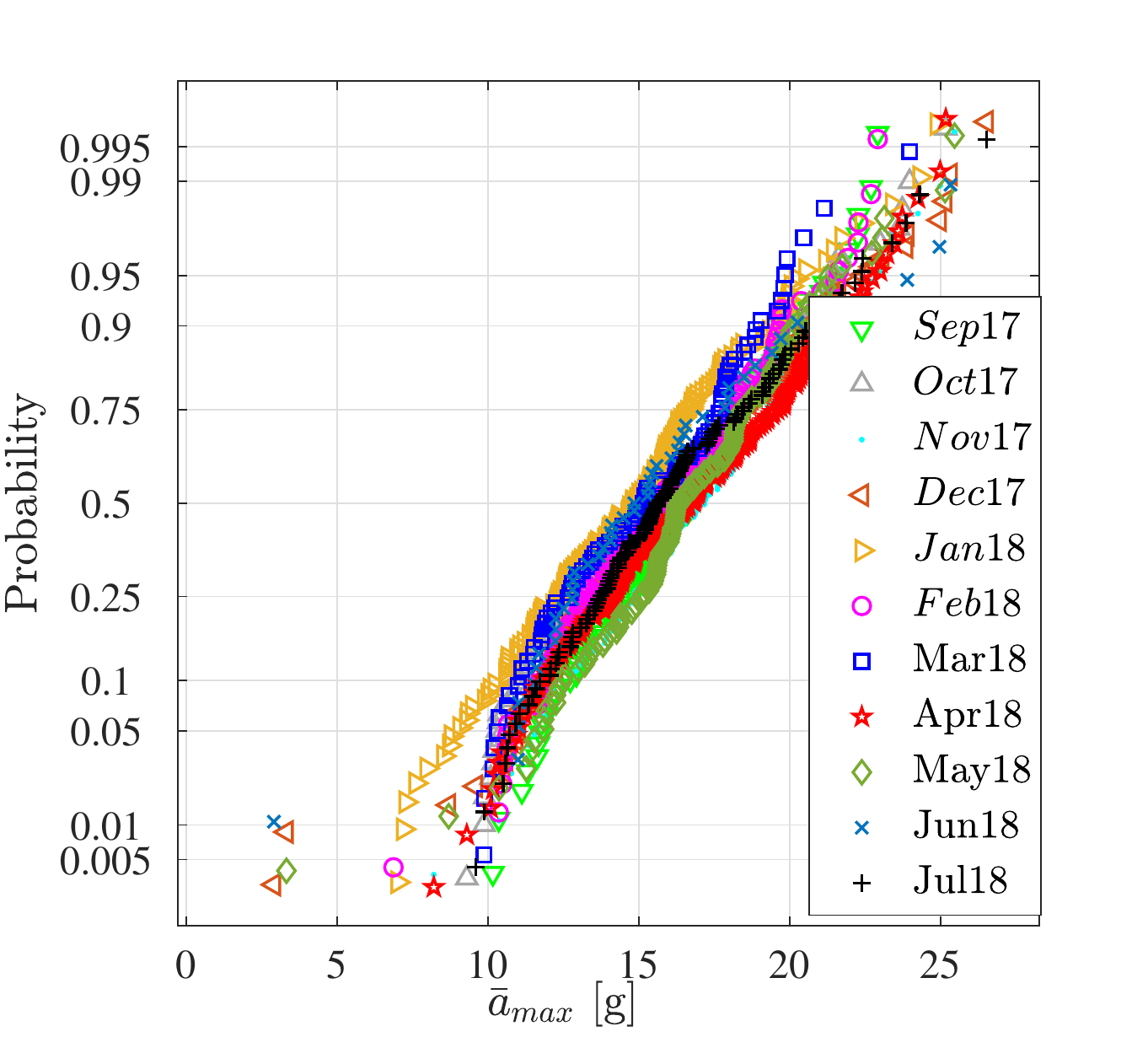}%
		\label{}}
	\caption{Statistical characteristic of the averaged maximum acceleration; IC3 trains at locations A2 and A4.}
	\label{fig:maxaccA2A4}
	\end{figure}

\begin{figure}[tbp] 
	\centering
	\subfloat[A7 ]{\includegraphics[width=0.48\columnwidth]{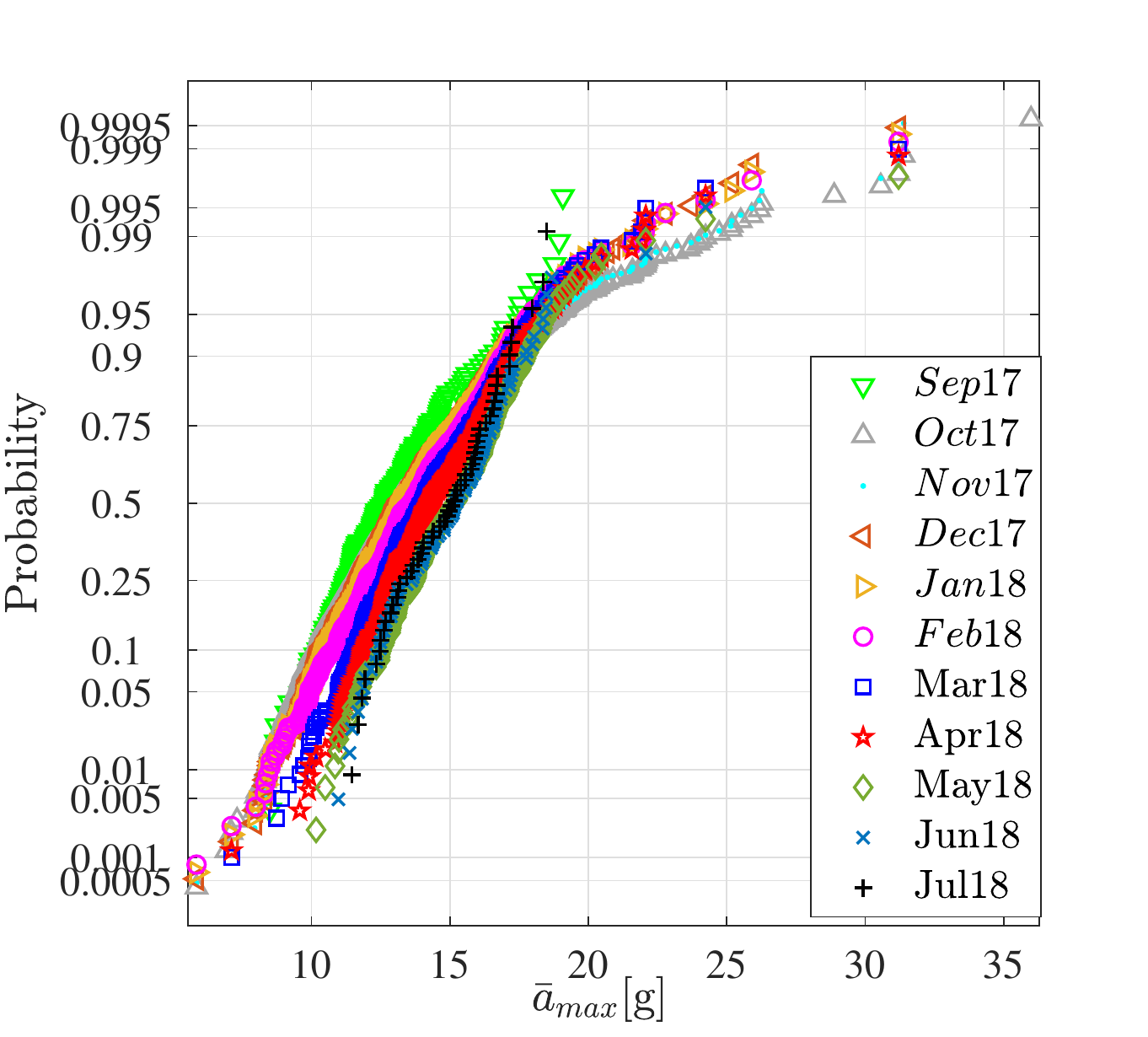}%
		\label{}}
	\hspace{0.05cm}	
	\subfloat[ A11]{\includegraphics[width=0.48\columnwidth]{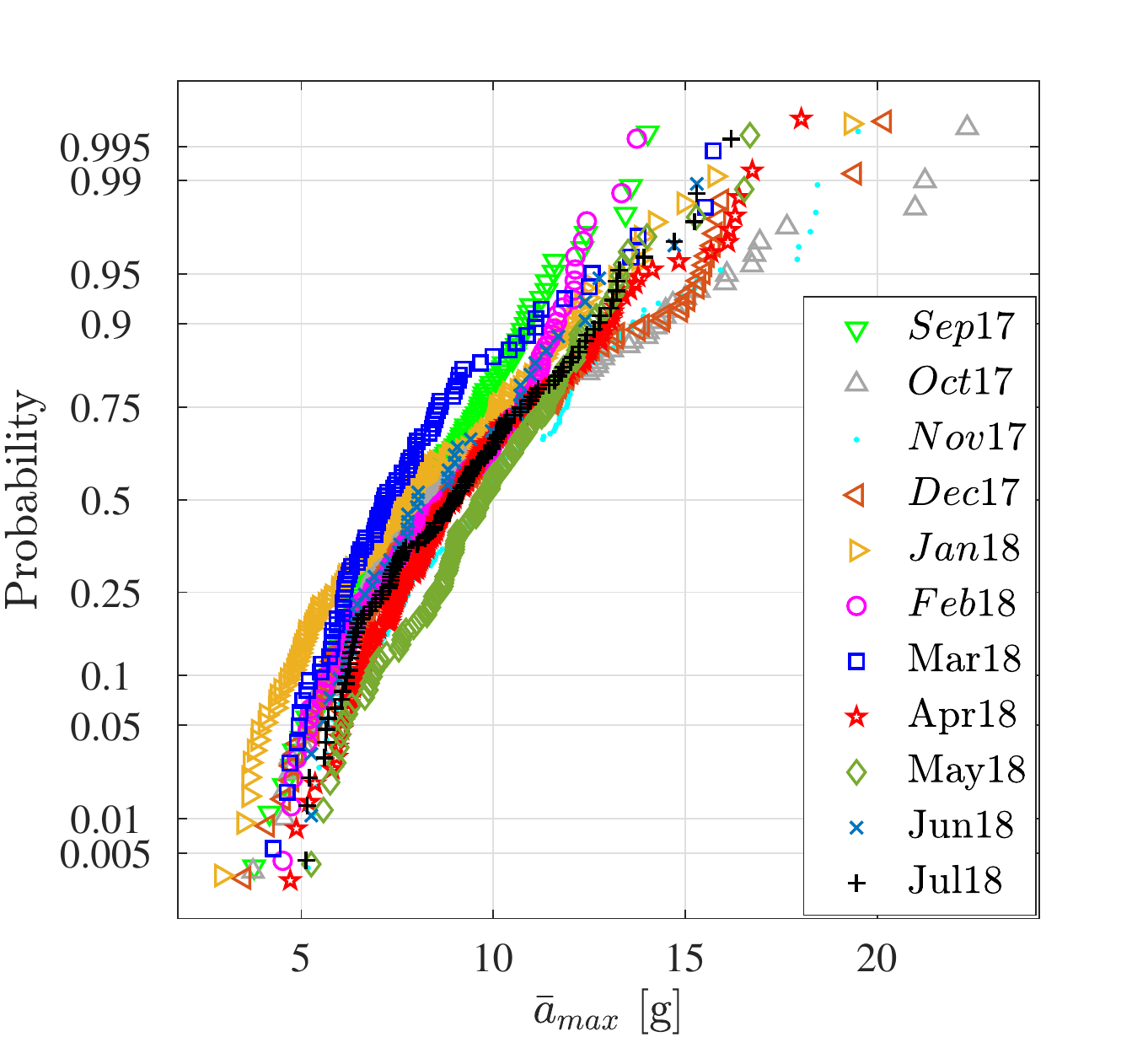}%
		\label{}}
	\caption{Statistical characteristic of the averaged maximum acceleration; IC3 trains at locations A7 and A11.}
	\label{fig:maxaccA7A11}
	\end{figure}

For a given train with $N_B$ bogies, the maximum track acceleration $a_{max}$ caused by the passage of each bogie is obtained. The mean value of the maximum accelerations $ \bar{a}_{max}$ is then calculated for each train
\begin{equation}
    \bar{a}_{max}=\frac{1}{N_B} \sum_{n=1}^{N_B} a_{max}(n).
\end{equation}
At each location along the turnout (A2, A4, A7 and A11), values of $\bar{a}_{max}$ are extracted from the acceleration signals induced by IC3 trains, with the speed ranging from 155 to 160 km/h, over monthly periods. Probability plots for the obtained values exhibit a similar behavior over the total period, as shown in Figs.~\ref{fig:maxaccA2A4}-\ref{fig:maxaccA7A11}. In fact, there is a small difference between the probability distributions in relation to the mean value and variance. The obtained results confirm that the effect of variations in the axle loads of the considered IC3 trains is statistically negligible.

\subsection{Temperature effect}
The 11 months period September 2017 -- July 2018 is considered for analyzing the effect of ambient temperature on the estimated frequencies associated with the behavior of the railpad layer. The considered period covers a wide range of temperature variation from \ang{-5}C to \ang{30}C. Hourly temperature data collected at the \AA{rslev} weather station (located 30 km South-East of Tommerup station) is provided by the Danish Meteorological Institute for the whole period. 
To assess the relation between the ambient temperature and the second track resonance frequency, variations of the estimated frequencies in the temperature domain are investigated. The whole temperature range from \ang{-5}C to \ang{30}C is divided into 18 segments. The following steps are then executed to obtain temperature-frequency data points:
(i) utilizing the hourly temperature data and the train passage time information provided by the track-side measurement system, all the acceleration signals are categorized into 18 groups according to the temperature segment in which they are measured; (ii) by applying the proposed method (stages 1 and 2) described in Section \ref{sec:method_railpad} to the acceleration signals in each group, a vector containing estimations of the second track resonance frequency is obtained; (iii) the median is taken over all frequency estimations in each vector; (iv) for each group, the calculated median and the mean value of the corresponding temperature segment are considered as a temperature-frequency data point.

\begin{figure}[tbp] 
 	\centering
 	\subfloat[A2 location]{\includegraphics[width=0.49\columnwidth]{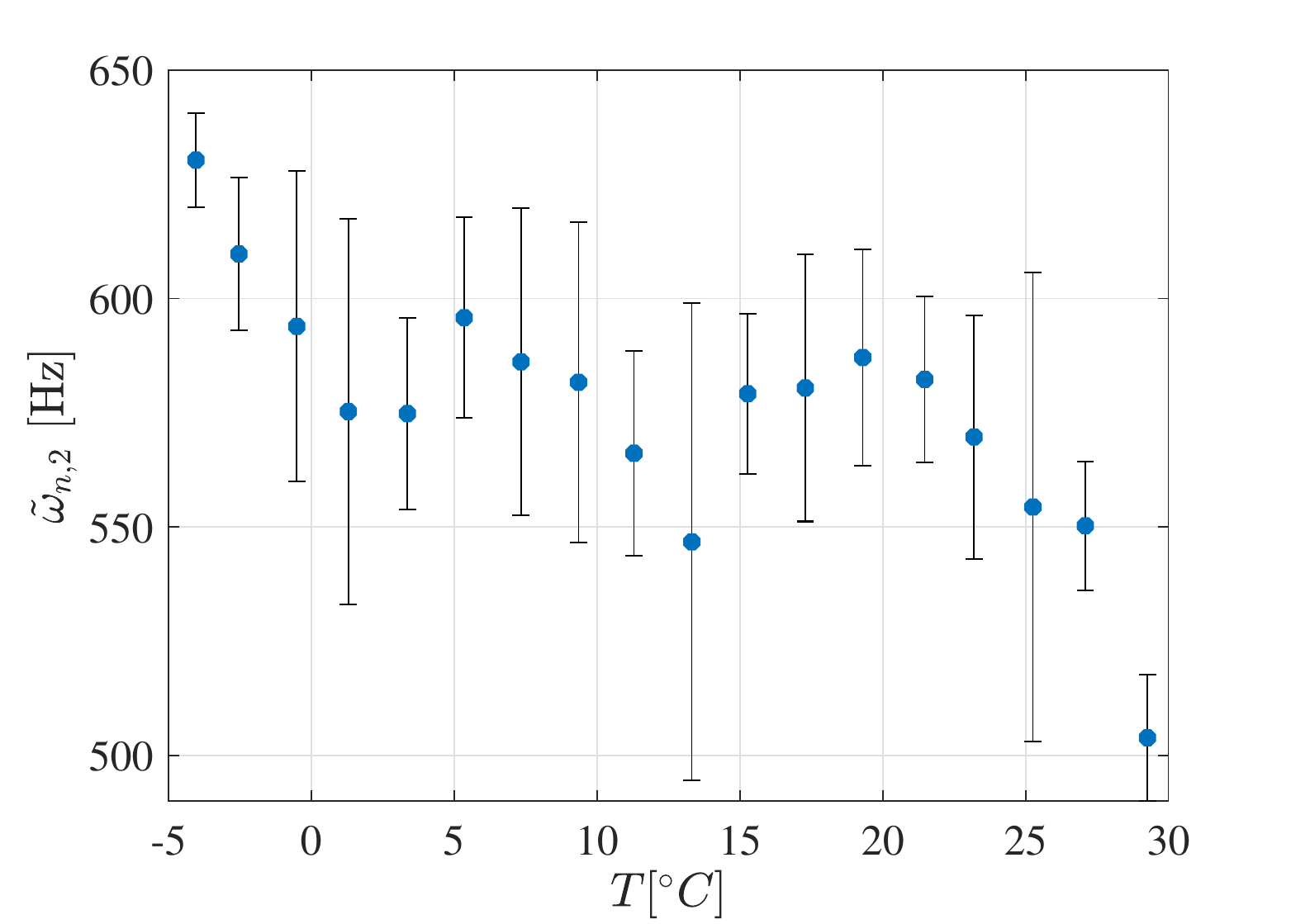}%
 		\label{}}
 	\hspace{0.05cm}	
 	\subfloat[ A4 location]{\includegraphics[width=0.49\columnwidth]{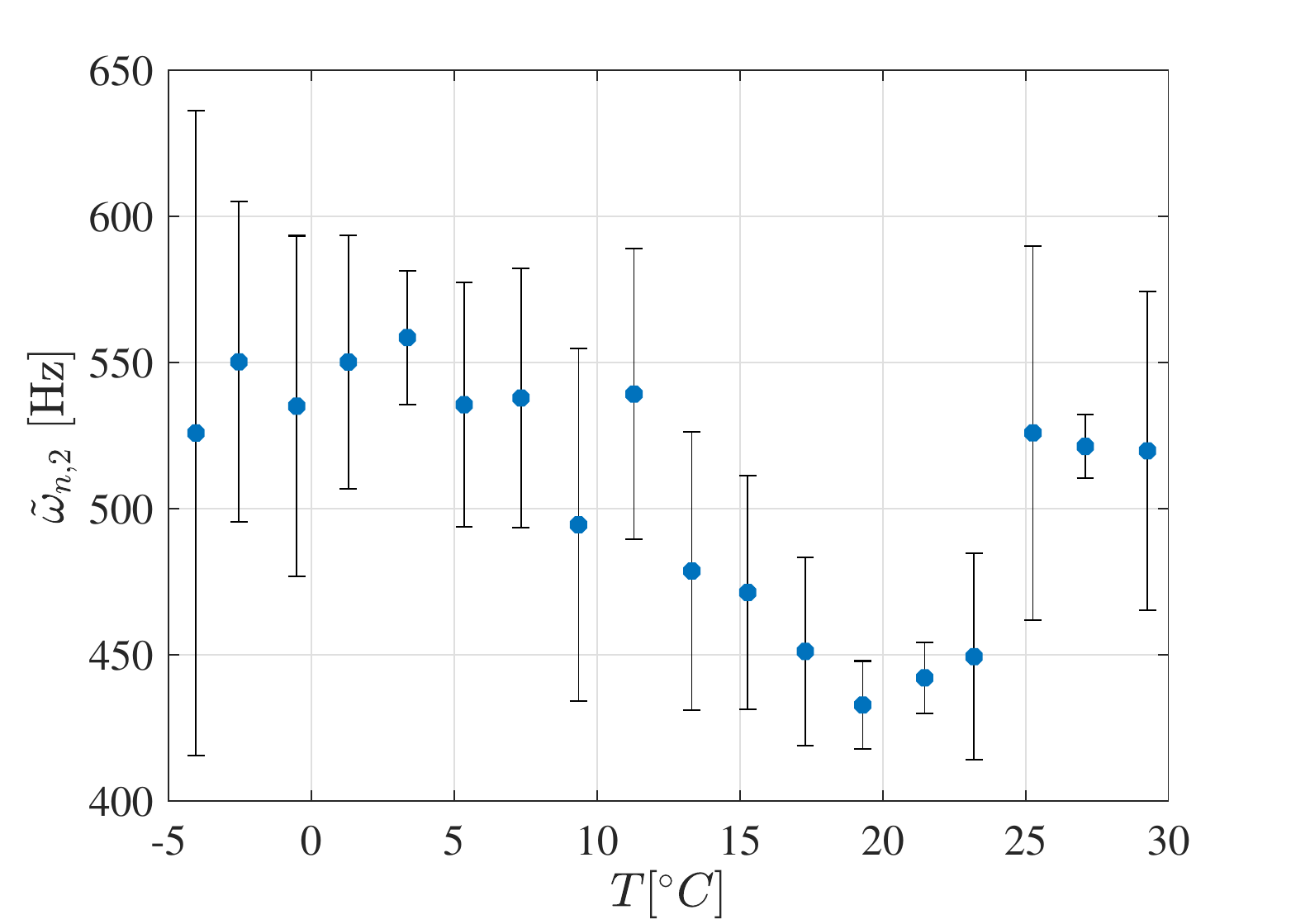}%
 		\label{}}
 	\caption{Median of the estimated second track resonance frequency with respect to the ambient	temperature at locations A2 and A4.}
	\label{fig:tempA2A4}
\end{figure}

\begin{figure}[tbp] 
 	\centering
 	\subfloat[A7 location]{\includegraphics[width=0.49\columnwidth]{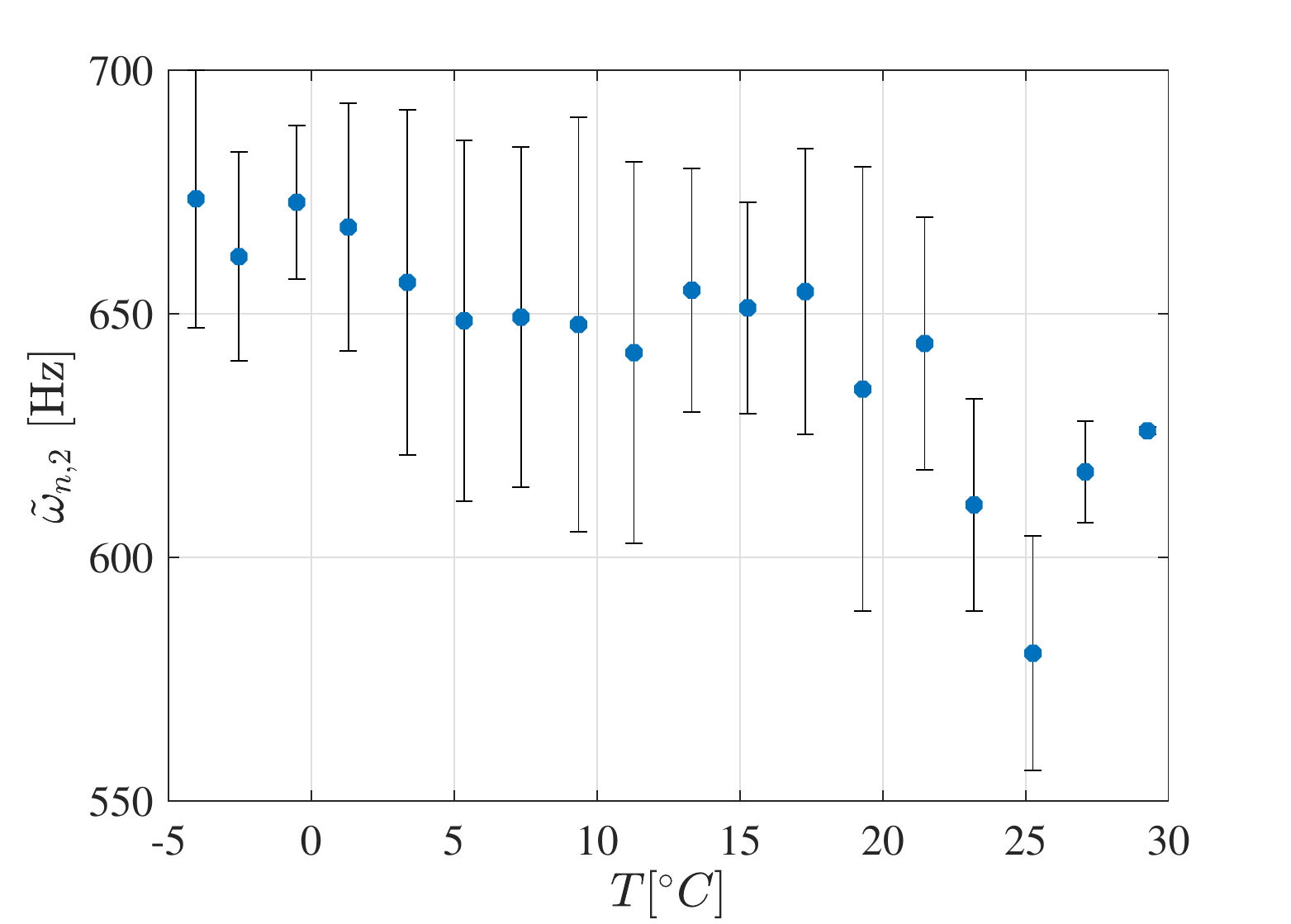}%
 		\label{}}
 	\hspace{0.05cm}	
 	\subfloat[ A11 location]{\includegraphics[width=0.49\columnwidth]{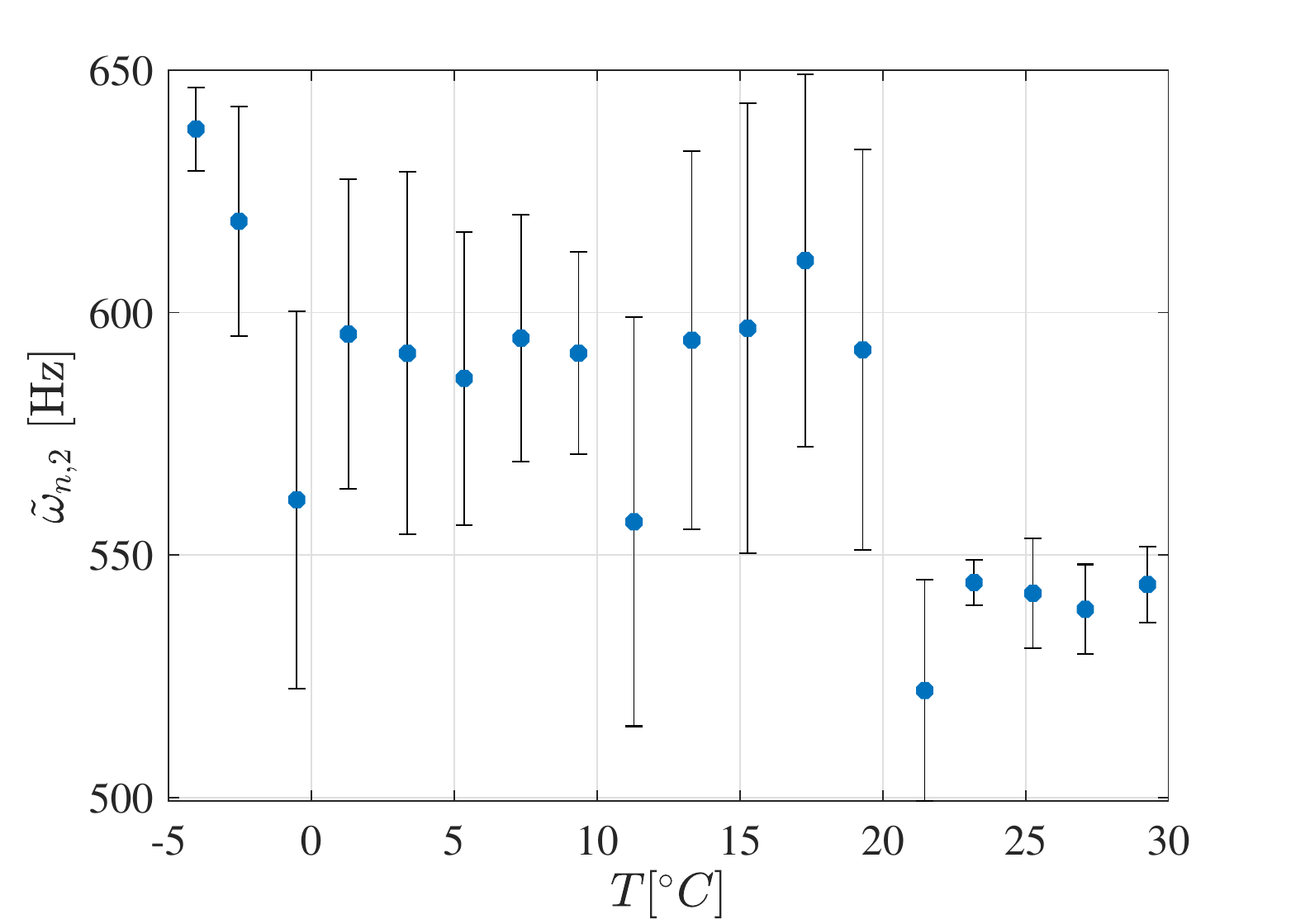}%
 		\label{}}
 	\caption{Median of the estimated second track resonance frequency with respect to the ambient	temperature at locations A7 and A11.}
	\label{fig:tempA7A11}
\end{figure}

\begin{figure}[tbp]
	\begin{center}
		\includegraphics[width=\linewidth]{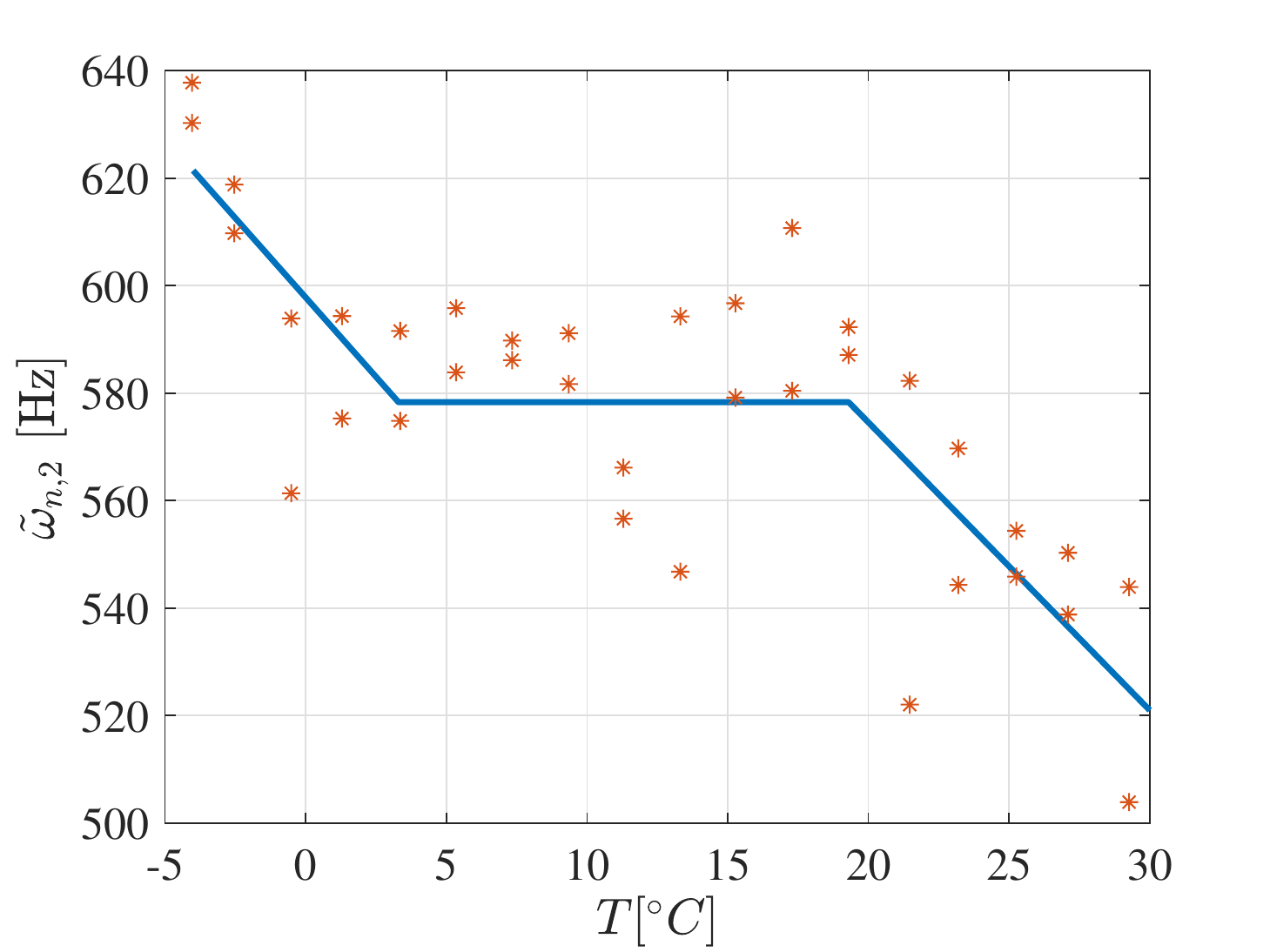}    
	\caption{Piecewise linear regression model fitted to the median of the estimated track resonances.} 
		\label{fig:piecewise}
	\end{center}
\end{figure}

Figures~\ref{fig:tempA2A4}-\ref{fig:tempA7A11} show the medians of the estimated track resonances and their median absolute deviations (MAD) plotted with respect to ambient temperature, for locations A2, A4, A7 and A11. It can be seen that the estimated resonance frequency decreases with an increase in temperature, at all the locations except A4. Although railpads of the same type are installed at all locations along the turnout, A2 and A11 are the only locations at which a similar frequency (y-axis) range is seen. The different frequency range observed at locations A4 and A7 is due to the different dynamic behavior at these locations caused by variations in the track geometry along the turnout, i.e., change in sleeper length at the crossing panel (A7) and rail discontinuity at the switch blades area (A4)~\cite{wan2014analysis}. The data points obtained at locations A2 and A11 are combined to derive a robust model representing the behavior of the second track resonance frequency as a function of temperature. 
Figure~\ref{fig:piecewise} shows the combined data points and a piecewise linear regression model fitted to them. This type of regression is chosen to improve the robustness of the temperature-frequency model. The reason is that the temperature data (x-axis) is hourly measured with highly accurate temperature sensors whereas the estimated frequency data (y-axis) is affected by different sources of uncertainty such as small load variations, rail corrugations, wheel defects and measurement noise.
The mathematical representation of the piecewise linear regression model employed to approximate the relation between the ambient temperature and the second track resonance frequency  for $T \in [-5~30] ^{\circ}C$ is given by,
\begin{equation}
    \mathcal{M}_T=
    \begin{cases}
      (-5.9\pm1.3)T+(597.8\pm10.4), &  T\leq3.3\\
    578.8\pm9.4, & 3.3<T\leq19.3\\
     (-5.3\pm2.4)T+(681.7\pm6.2) & T>19.3
    \end{cases}
    \label{eq:model_railpad}
\end{equation}

\begin{figure}[tbp] 
 	\centering
 	\subfloat[A4 location]{\includegraphics[width=0.49\columnwidth]{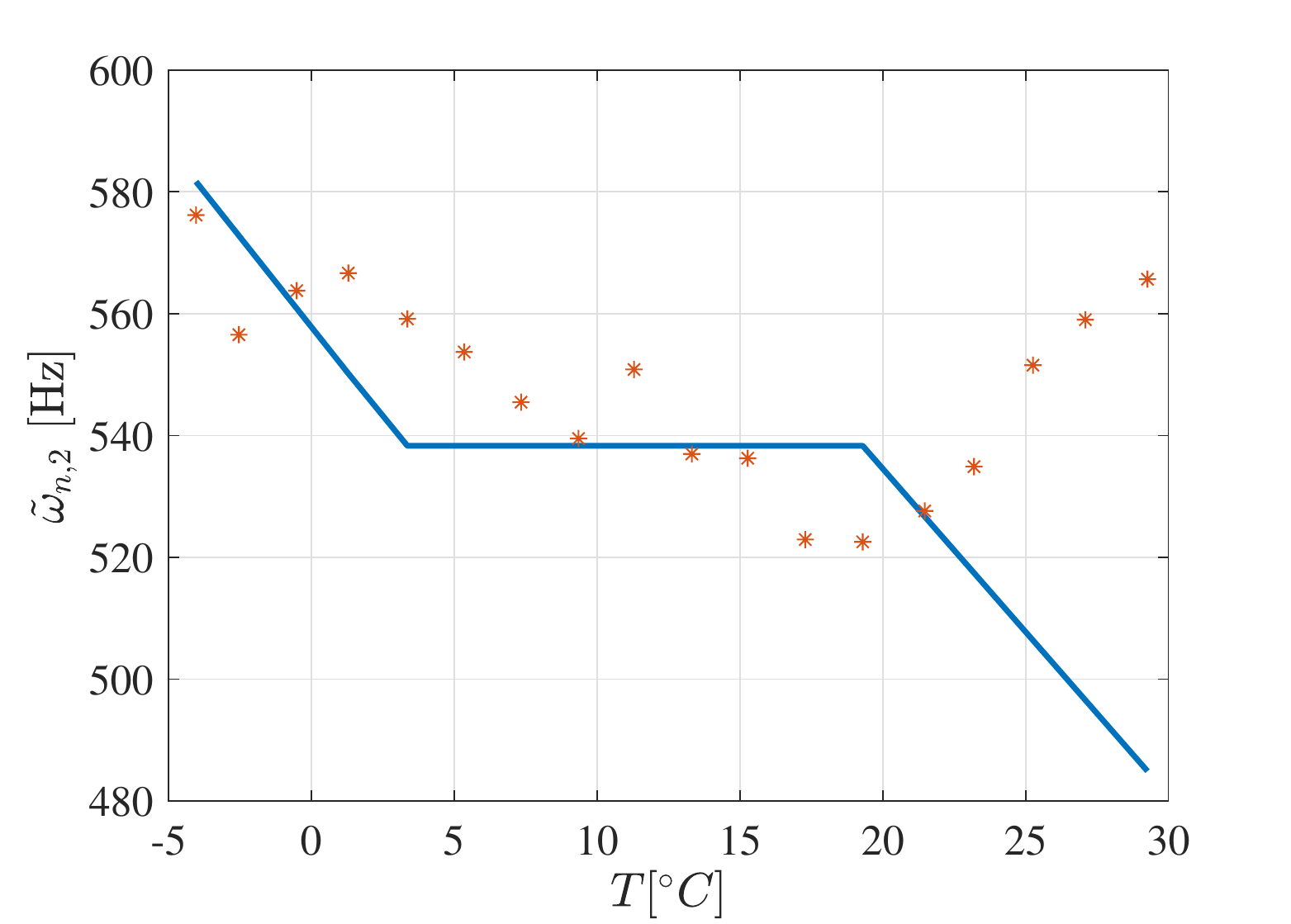}%
 		\label{}}
 	\hspace{0.05cm}	
 	\subfloat[ A7 location]{\includegraphics[width=0.49\columnwidth]{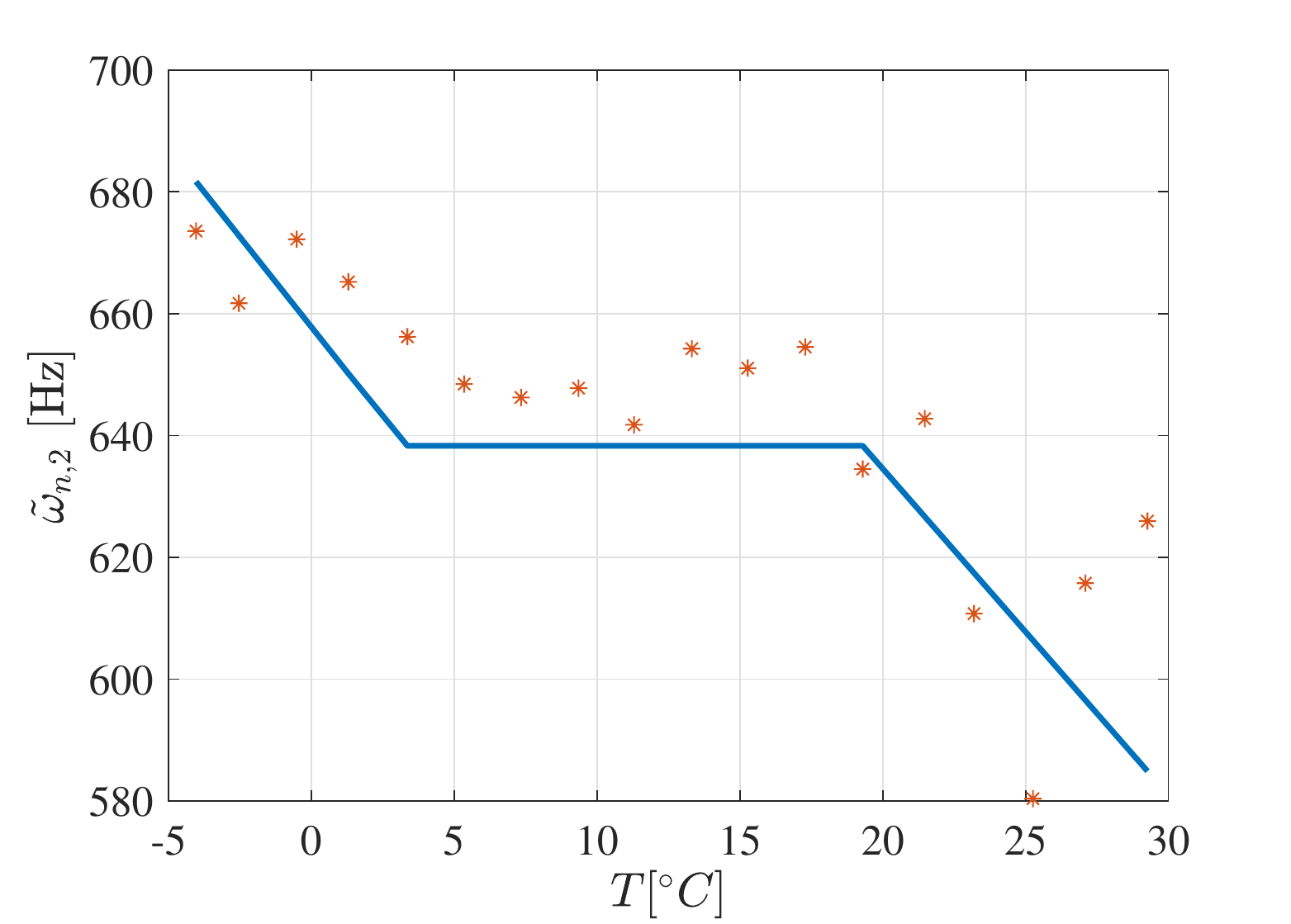}%
 		\label{}}
 	\caption{Piecewise linear regression model compared with the median of the estimated second track resonance frequency with respect to the ambient temperature at locations A4 and~A7.}
	\label{fig:tempmodelA4A7}
\end{figure}

As mentioned earlier, the same type of railpad is used along the entire turnout. A similar temperature-frequency behavior is therefore expected for the railpads at all the locations. In order to find regression models describing the temperature dependence of the second track resonance at locations A4 and A7, the piecewise linear regression model in \eqref{eq:model_railpad} is shifted along the y-axis ($\tilde{\omega}_{n,2}$ axis). Figure \ref{fig:tempmodelA4A7} shows the median of the second track resonance frequency with respect to temperature, and the shifted piecewise linear regression model for locations A4 and A7.
As it can be seen, the model is capable to properly describe the relation between the second track resonance and the ambient temperature, at location A7. However, the temperature-frequency data points obtained for location A4 exhibit a different behavior, particularly for temperatures greater than $20 ^{\circ}C$. This different behavior can be due to a defect/malfunction or deterioration in the quality of the railpad at this location. However, the root cause of this behavior has not been identified since it was not possible to dismantle the track and examine the quality of the in-service railpad at this location.

The temperature-frequency piecewise linear regression model is used in this study to isolate temperature-related variations of the second track resonance frequency from the variations caused by railpad defects or deterioration (i.e., aging effect).
\section{Monitoring tool design} \label{sec:monitor_railpad}
In order to design a tool for monitoring the condition of the railpads and detect the deterioration process occurring over time, a detection signal in which the temperature effect is minimized must be defined. For this purpose, the following steps are carried out
\begin{enumerate}[label=(\roman*)]
    \item The hourly temperature data and time information provided by the track-side measurement system are used to assign a temperature to each estimation of resonance frequency $(\hat{\omega}_{n,2})_i$ extracted from a recorded acceleration signal $a_i$ by using the method discussed in Section~\ref{sec:method_railpad}.
    \item The assigned temperature is utilized as input to the piecewise linear regression model to predict the second track resonance frequency, $(\hat{\omega}^{\mathcal{M}}_{n,2})_i$.
    \item  Steps (i) and (ii) are repeated for all acceleration signals recorded over monthly periods to obtain the frequency vectors $\hat{\boldsymbol\omega}_{n,2}$ and  $\hat{\boldsymbol\omega}^{\mathcal{M}}_{n,2}$ (see Fig.~\ref{fig:railpadmonitoring}).
    \item The detection signal is then defined as $\mathbf{r} = \hat{\boldsymbol\omega}_{n,2} - \hat{\boldsymbol\omega}^{\mathcal{M}}_{n,2}$.
\end{enumerate}

\subsection{Statistical characterization}
In the process of designing a statistical change detection algorithm the distribution that well approximate the statistical behavior of the residual $\mathbf{r}$ must be determined. Suitability of different distributions has been examined. Among those, three distributions which can better represent the behaviour of the residual sequence, i.e. the Gaussian distribution ($\mathcal{N}$), the Weibull distribution ($\mathcal{W}$) and the Generalized Extreme Value distribution ($\mathcal{G}$), are compared to find the best fit to the data.

\begin{figure}[tbp] 
	\centering
	\subfloat[A2]{\includegraphics[width=0.48\columnwidth]{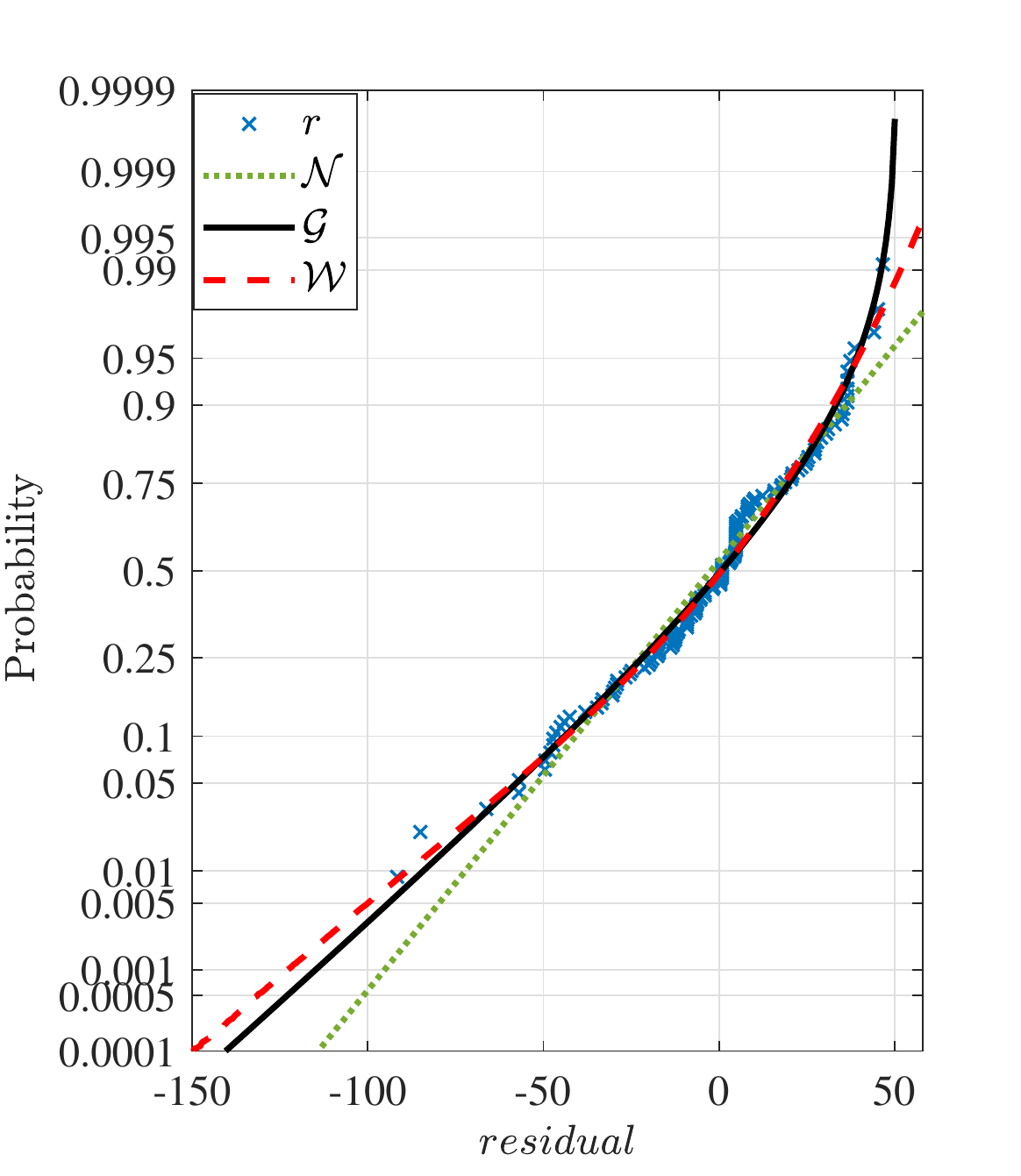}%
		\label{}}
	\hspace{0.05cm}	
	\subfloat[A4]{\includegraphics[width=0.48\columnwidth]{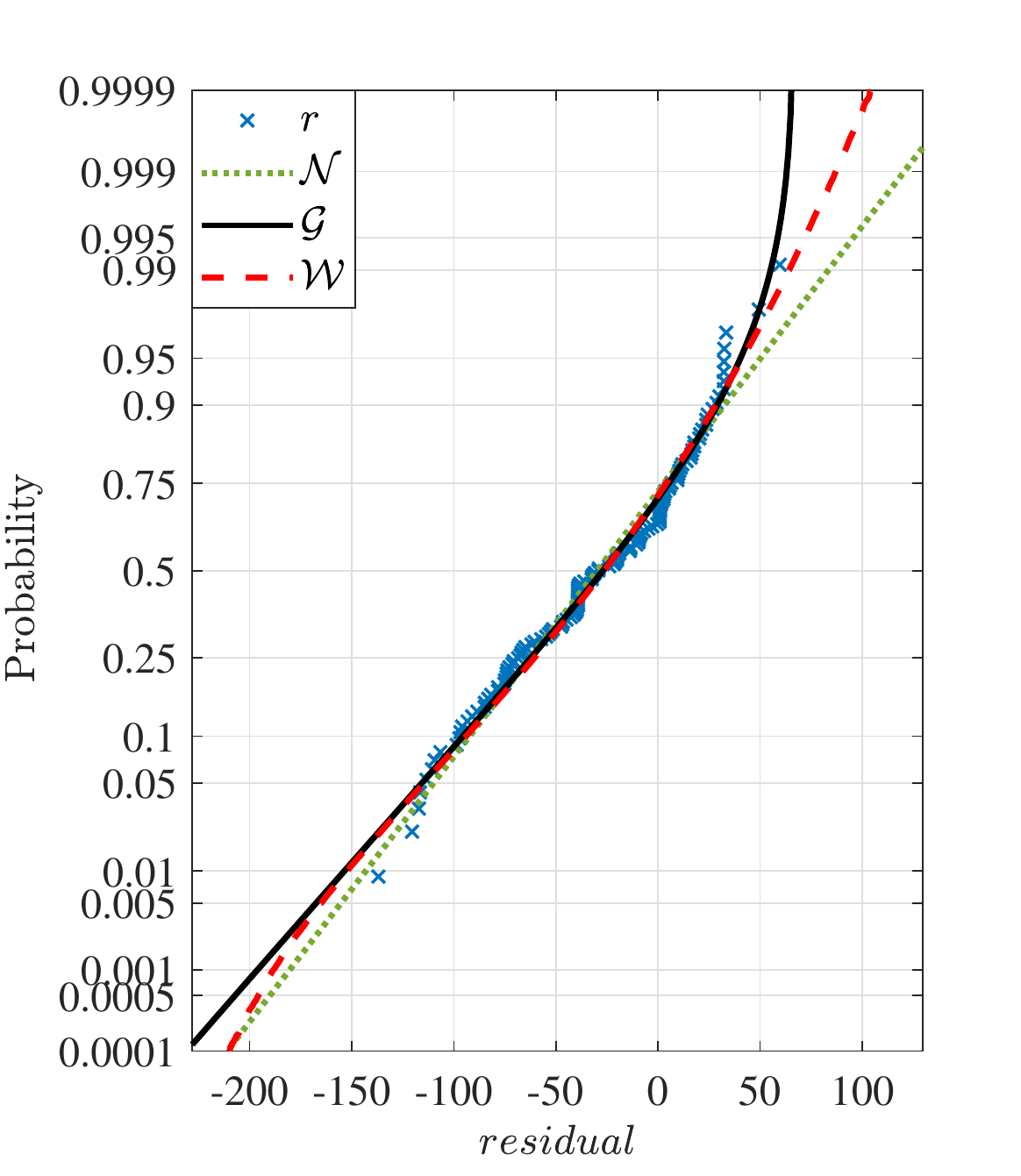}%
		\label{}}
	\caption{Probability plots for residual sequence at locations A2 and  A4. Weibull ($\mathcal{W}$), Gaussian ($\mathcal{N}$), and GEV ($\mathcal{G}$) distributions fitted to the data.}
	\label{fig:resA2A4}
	\end{figure}

\begin{figure}[tbp] 
	\centering
	\subfloat[A7]{\includegraphics[width=0.48\columnwidth]{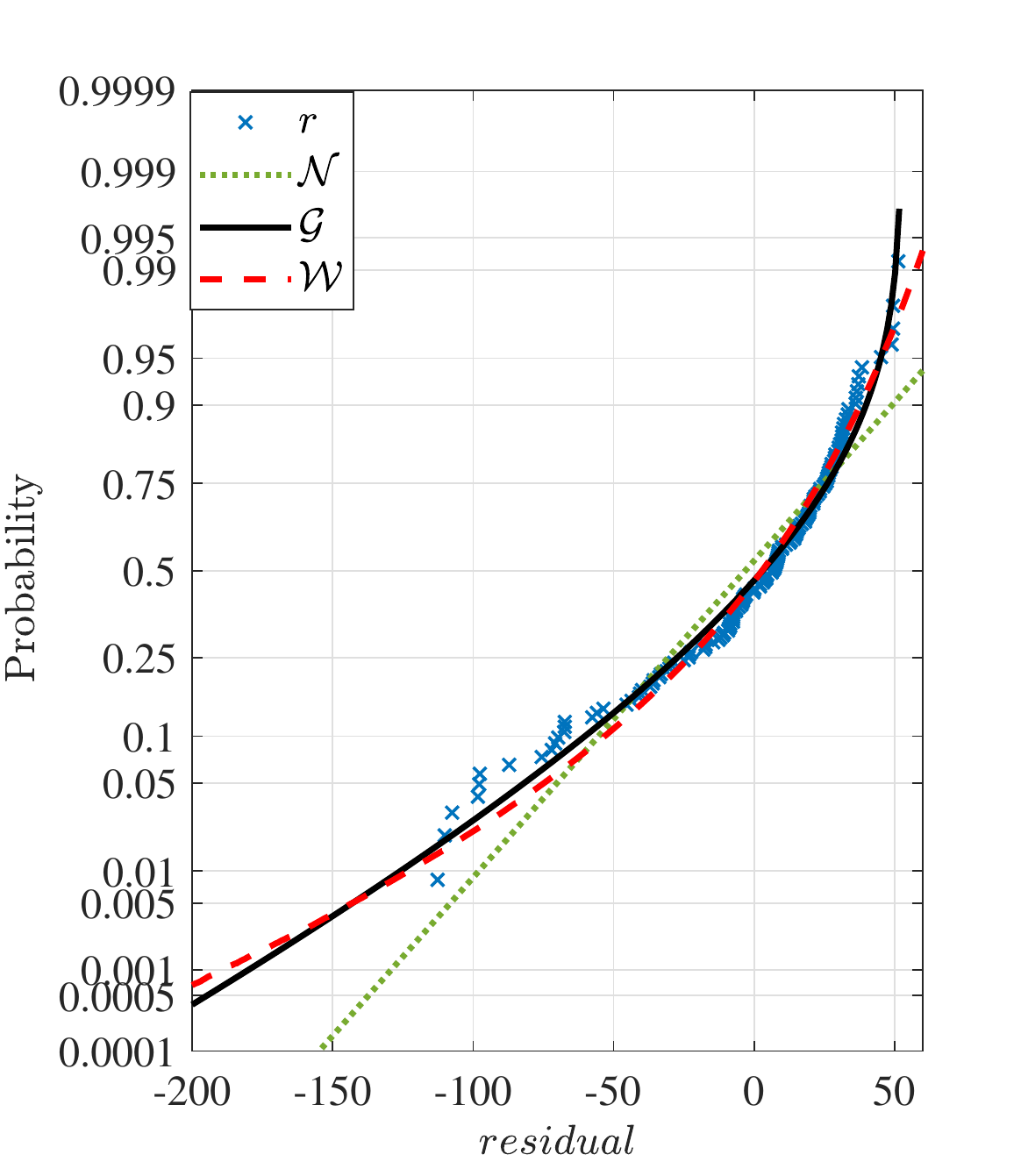}%
		\label{}}
	\hspace{0.05cm}	
	\subfloat[A11]{\includegraphics[width=0.48\columnwidth]{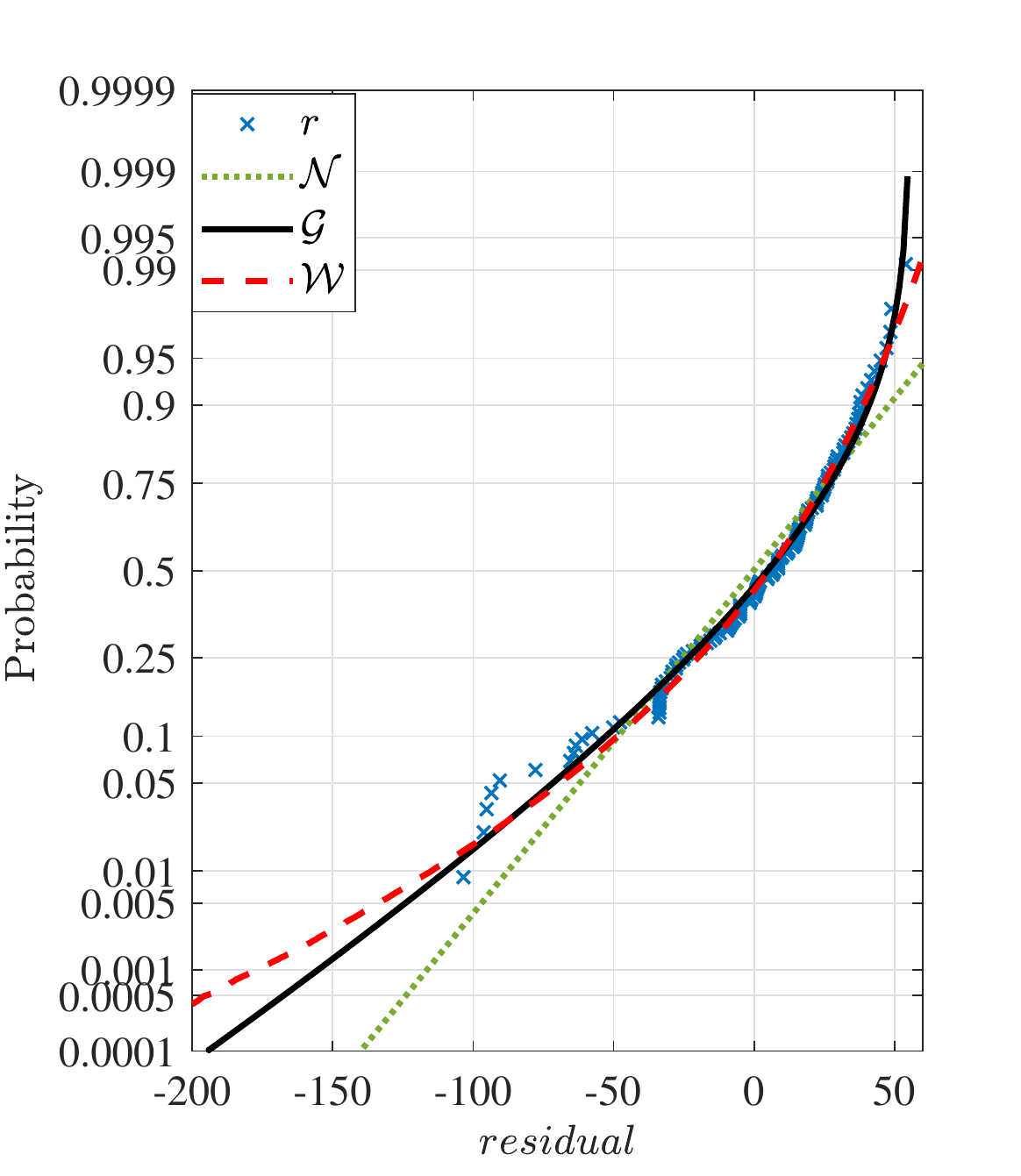}%
		\label{}}
	\caption{Probability plots for residual sequence at locations A7 and  A11. Weibull ($\mathcal{W}$), Gaussian ($\mathcal{N}$), and GEV ($\mathcal{G}$) distributions fitted to the data.}
	\label{fig:resA7A11}
	\end{figure}

Figures~\ref{fig:resA2A4}-\ref{fig:resA7A11} show the probability plots of the residual sequences obtained for locations A2, A4, A7 and A11 in September 2017 and the three types of distributions fitted to them. It can be seen that the generalized extreme value (GEV) distribution better characterizes the statistical behavior of the residual sequences. Parameters of the obtained distributions and their goodness of fit calculated using the Kolmogorov-Smirnov (K-S) test are given in Table \ref{tab:tablefitdist}, which also confirm that the GEV distribution is the best fit to the data at all considered locations.
\begin{table*}[tbp]
\centering
\caption{Parameters of the considered distributions and their corresponding $p$-value}
\label{tab:tablefitdist}
\begin{tabular}{ccccc}
\toprule
Location & residual & Distribution & Parameters &  $p$-value\\ 
\midrule 
& \multirow{3}{*}{$r$} &$\mathcal{G}$ & $\xi=-0.48$, $\sigma=28.71$, $\mu=-8.97$ &0.51 \\
 $A_2$ &  & $\mathcal{W}$ & $\alpha=10.24$, $\beta=237.10$, $\mu=-228.04$  & 0.13 \\
& & $\mathcal{N}$ & $\sigma=26.65$, $\mu= -2.24$ & 0.19 \\
	\midrule 
& \multirow{3}{*}{$r$} &$\mathcal{G}$ & $\xi=-0.41$, $\sigma=45.24$, $\mu=-42.30$ &   0.32\\
 $A_4$ &  & $\mathcal{W}$ & $\alpha=5.97$, $\beta=233.56$, $\mu=-246.05$ & 0.22 \\
& & $\mathcal{N}$ & $\sigma=43.15$, $\mu= -29.84$ &0.13 \\
	\midrule
	& \multirow{3}{*}{$r$} &$\mathcal{G}$ & $\xi=-0.62$, $\sigma= 38.65$, $\mu= -9.62 $ & 0.97 \\
 $A_7$ &  & $\mathcal{W}$ &  $\alpha=49.72$, $\beta=1304.84$, $\mu=-1292.40$  & 0.73 \\
& & $\mathcal{N}$ & $\sigma=36.14$, $\mu=-3.11$ & 0.02 \\
	\midrule
		& \multirow{3}{*}{$r$} &$\mathcal{G}$ & $\xi=-0.57$, $\sigma=35.77$, $\mu=-7.04$ & 0.74 \\
 $A_{11}$ &  & $\mathcal{W}$ &  $\alpha=39.98$, $\beta=1014.71$, $\mu=-1000.33$  & 0.70\\
& & $\mathcal{N}$ & $\sigma=33.32$, $\mu= -0.18$ &  0.12 \\
\bottomrule
\end{tabular}
\end{table*}

The probability density function (PDF) of the GEV distribution chosen for describing the behavior of the residual sequence is given by,
\begin{equation}
\mathcal{G}(r) =
\frac{1}{\sigma} \exp\left[-\left(1+\xi\frac{r-\mu}{\sigma}\right)\right]^{-\frac{1}{\xi}} \left(1+\xi\frac{r-\mu}{\sigma}\right)^{-1-\frac{1}{\xi}}, \label{eq:gev}
\end{equation}
defined on the set $\left\{r \, : \, 1+\xi(r-\mu)/\sigma > 0 \right\}$, where $\mu$, $\sigma$ and $\xi\neq 0$ are the location, scale and shape parameters, respectively. There are three types of GEV distribution known as Type I (Gumbel), Type II (Frechet) and Type III (Reversed Weibull). The shape parameter determines the type of the GEV distribution. $\xi=0$, $\xi>0$ and $\xi<0$ correspond to Type I, Type II and Type III, respectively. According to the obtained values of the shape parameter presented in Table \ref{tab:tablefitdist}, the GEV distribution representing the behavior of the residual sequence is of Type III.

\begin{figure}[tbp] 
 	\centering
 	\subfloat[A2 location]{\includegraphics[width=0.49\columnwidth]{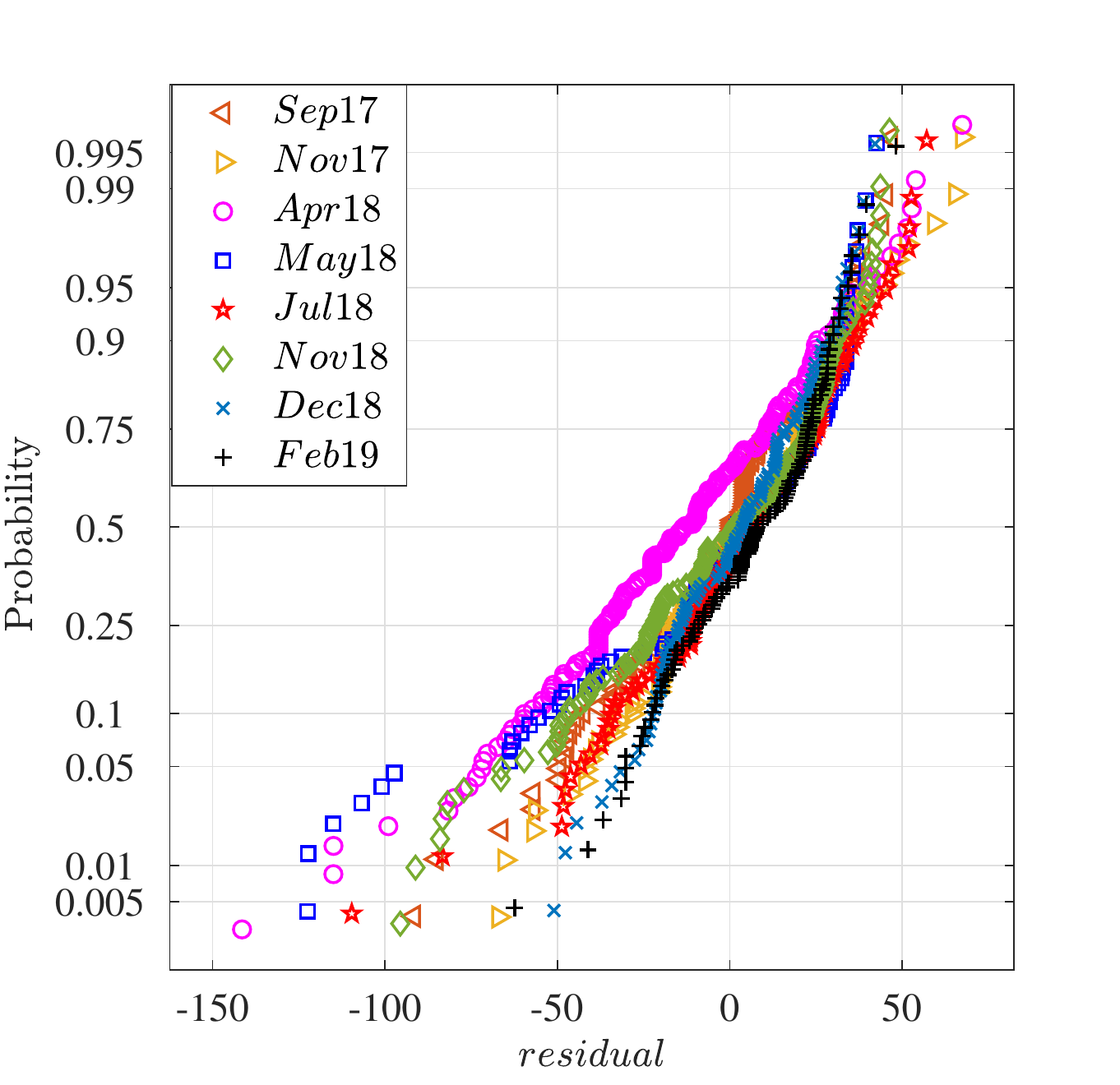}%
 		\label{}}
 	\hspace{0.05cm}	
 	\subfloat[ A4 location]{\includegraphics[width=0.49\columnwidth]{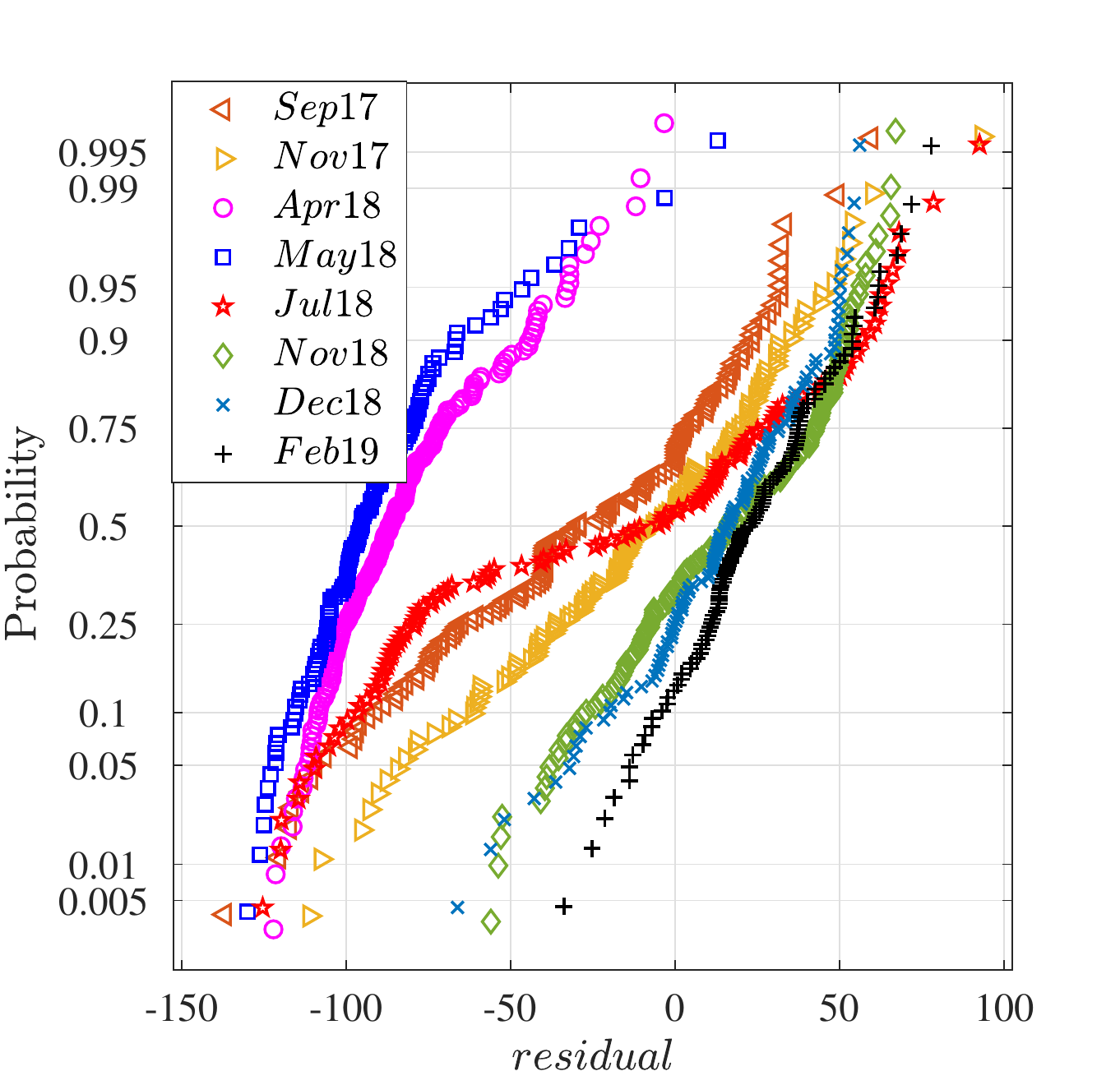}%
 		\label{}}
 	\caption{Probability plot of the residual sequences obtained at A2 and A4 location}
	\label{fig:residualprobA2A4} 
\end{figure}

Figure~\ref{fig:residualprobA2A4} shows the probability plots of the residual sequences calculated by considering the track acceleration data collected at locations A2 and A4 over a period of 18 months. For location A2, the obtained probability plots exhibit no significant variation over the considered period, indicating that the stiffness of the railpad at this location did not vary noticeably after isolating the temperature effect. A different behavior is observed at location A4 where the probability plot of the residual sequence  varies significantly over the considered period. This behavior implies that the change in the stiffness of the railpad layer at location A4 is caused by a factor different from temperature and load variations. One possible explanation for this behavior is that the railpad is worn or defective. As mentioned earlier in Section \ref{sec:temp}, this hypothesis has not been verified since it was not possible to disassemble the track.

In order to monitor the railpad quality over time, designing a statistical test for detecting changes in the residual sequence is required. The detection problem can be formulated as a parameter test,
\begin{align*}
    \mathcal{H}_0 &: \boldsymbol{\theta} = \boldsymbol{\theta}_0 \\
    \mathcal{H}_1 &: \boldsymbol{\theta} \ne \boldsymbol{\theta}_0
\end{align*}
where $\boldsymbol{\theta} = [\mu, \sigma, \xi]^\mathrm{T}$ is the vector of the GEV parameters and $\boldsymbol{\theta}_0$ is its known instance under the $\mathcal{H}_0$ hypothesis. Since the value of $\boldsymbol{\theta}$ under$\mathcal{H}_1$ is unknown
the generalized likelihood ratio test (GLRT) is employed to design a detector able to discriminate between the two competing hypothesis. 

To calculate the threshold for the change detection that results in the desired probability of false alarm, theoretical requirements of independent and identically distributed (iid) samples of the detection signal $\mathbf{r}$ should be investigated. In practical cases, when the residual sequence is non-iid, the actual distribution of the test statistic signal under $\mathcal{H}_0$ is employed to find a threshold value that provides the desired probability of false alarms. This was shown e.g., in \cite{blanke2016diagnosis,galeazzi2013early}.

\subsection{$\mathcal{G}$-GLRT detector design}
Considering the residual sequence as a realization of random GEV ``type III'' process, the distribution of $N$ iid samples is described by,
\begin{align}
\mathcal{G}(\mathbf{r};\boldsymbol{\theta}) &= \left(\frac{1}{\sigma}\right)^N \prod_{i=1}^{N}  \exp\left(-\left(1+\xi\left(\frac{r_i-\mu}{\sigma}\right)\right)^{-\frac{1}{\xi}}\right) \nonumber \\
&\phantom{=}\; \times\prod_{i=1}^{N} \left(1+\xi\left(\frac{r_i-\mu}{\sigma}\right)\right)^{-1-\frac{1}{\xi}}, \label{eq:GEVN}
\end{align}
in which $\mathbf{r}=[r(k-N),...,r(k)]^{\mathrm{T}}$, and $N$ is the number of identical IC3 trains, with the speed ranging from 155 to 160$\,\mathrm{km}/\mathrm{h}$, passing the turnout over monthly periods.

The GLRT detector decides $\mathcal{H}_1$ if
\begin{equation}
L_G(\mathbf{r})=\frac{\mathcal{G}(\mathbf{r};\boldsymbol{\hat{\theta}},\mathcal{H}_1)}{\mathcal{G}(\mathbf{r};\boldsymbol{{\theta}_0},\mathcal{H}_0)} >\gamma , \label{eq:GLRT}
\end{equation} 
where $\boldsymbol{\hat{\theta}}$ denotes the maximum likelihood estimation (MLE) of the GEV parameters under the hypothesis $\mathcal{H}_1$ and $\gamma$ is the threshold calculated to provide the desired probability of false alarms. In order to compute $L_G(\mathbf{r}$) in \eqref{eq:GLRT}, the vector of the unknown parameters $\boldsymbol{{\theta}}=[{\mu},{\sigma},{\xi}]^\mathrm{T}$ must be determined first. This is performed by maximizing $\mathcal{G}(\mathbf{r};\boldsymbol{{\theta}},\mathcal{H}_1)$ with respect to $\boldsymbol{{\theta}}$. The log-likelihood for the GEV parameters is~\cite{coles2001introduction},
\begin{align}
\ell(\boldsymbol{{\theta}})&=-N\ln\sigma-\left(1+\frac{1}{\xi}\right)\sum_{i=1}^{N}\ln\left[1+\xi\left(\frac{r_i-\mu}{\sigma}\right)\right] \nonumber \\ 
&\phantom{=}\; - \sum_{i=1}^{N} \left[1+\xi\left(\frac{r_i-\mu}{\sigma}\right)\right]^{-\frac{1}{\xi}},
  \label{eq:loglik}
\end{align} 
when $\xi \neq 0$ and,
\begin{equation}
 \left[1+\xi\left(\frac{r_i-\mu}{\sigma}\right)\right]>0. \nonumber
\end{equation} 
The MLE of $\boldsymbol{{\theta}}$ is obtained from \eqref{eq:loglik} by employing standard numerical optimization algorithms, since an analytical solution does not exist.

Substituting \eqref{eq:GEVN} into \eqref{eq:GLRT}, the natural logarithm is taken of both sides of the resultant equation, and vector of parameters $\boldsymbol{{\theta}}$ is then replaced with $\boldsymbol{\hat{\theta}}$, in order to obtain the explicit form of the detector. For the $k_w$-th time window containing the residual sequence  calculated over a one-month period the test statistic $g(k_w)$ is obtained as,
\begin{align}
g(k_w) &= N \ln\left(\frac{{\sigma_0}}{{\hat\sigma}}\right)-\left(1+\frac{1}{\hat\xi}\right)\sum_{i=1}^{N}\ln\left[1+\hat\xi\left(\frac{r_i-\hat\mu}{\hat\sigma}\right)\right] \nonumber \\
&\phantom{=}\;
+(1+\frac{1}{\xi_0})\sum_{i=1}^{N}\ln\left[1+\xi_0\left(\frac{r_i-\mu_0}{\sigma_0}\right)\right] \nonumber \\ 
&\phantom{=}\;
-\sum_{i=1}^{N} \left[1+\hat\xi\left(\frac{r_i-\hat\mu}{\hat\sigma}\right)\right]^{-\frac{1}{\hat\xi}} \nonumber \\
&\phantom{=}\;
+\sum_{i=1}^{N} \left[1+\xi_0\left(\frac{r_i-\mu_0}{\sigma_0}\right)\right]^{-\frac{1}{\xi_0}} > \gamma\prime \label{eq:g(k)},
\end{align}
where  $\gamma\prime$ is $\ln \gamma$.

According to the Neyman-Pearson theorem~\cite[Chapter 3]{kay1998fundamentals}, if the residual sequence is iid, then the threshold $\gamma$ given by the maximum probability of detection $P_{D}$ with a desired probability of false alarms $P_{FA}$ can be calculated as solution of the integral
\begin{equation}
P_{FA}=\int_{L_G(\mathbf{r})>{\gamma}} \mathcal{G}(\mathbf{r};\boldsymbol{\theta}_0,\mathcal{H}_0 ) \, \mathrm{d}\mathbf{r}. \label{eq:PFA}
\end{equation}
When the length of the sequence $\mathbf{r}$ is very large (i.e., $N \to \infty$), there exists an asymptotic result for the modified GLR test statistics $2L_G(\mathbf{r})$~\cite{kay1998fundamentals}. However, for the case studied in this paper the length of the residual sequence  is short and, therefore, the asymptotic results are not valid.

\section{Performance evaluation of the monitoring system} \label{sec:Impelement_railpad}
\subsection{$\mathcal{G}$-GLRT detector set-up and tuning}
The first step to get the $\mathcal{G}$-GLRT detector operational is to determine the vector $\boldsymbol{\theta}_0$. For the case considered in this study, $\boldsymbol{\theta}_0$ is not invariant in an absolute sense since it must be recomputed whenever worn railpads are replaced with new ones. The recalculation of $\boldsymbol{\theta}_0$ is carried out to ensure that the actual statistical characteristics of the residual sequence are captured after railpad replacement. In between the replacement actions, $\boldsymbol{\theta}_0$ is considered to be known and constant.

The monitoring action consists of two phases: 
\begin{enumerate}[label=(\roman*)]
    \item \textbf{Initialization}: For a given location along the turnout, after each railpad replacement event the residual sequence is obtained by using the track acceleration signals and the temperature data collected over a one-month period (estimation window, $M_E$). The MLE of the GEV distribution parameters under the hypothesis $\mathcal{H}_0$ is then computed for the obtained residual sequence.
    \item \textbf{Detection}: Once $\hat{\boldsymbol{\theta}}_0$ is estimated, the test statistic given in \eqref{eq:g(k)} is calculated utilizing the residual sequences obtained in monthly periods (detection window, $M_D$). This process is repeated iteratively until the next railpad replacement action takes place.
\end{enumerate}
The performance of the monitoring system depends on the selection of both the estimation window and the detection window. Proper selection of $M_E$ leads to a more accurate characterization of the residual sequence in the initialization phase. $M_D$ influences the test statistic results calculated during the detection phase. A lower number of fluctuations in the output of the $\mathcal{G}$-GLRT test statistic can be expected by choosing a large $M_D$; on the other hand a small $M_D$ allows a more timely detection. 
In this study, a one-month window size is found to be an appropriate choice which provides a good trade-off between the promptness and the accuracy of the monitoring tool.

\begin{figure}[tbp]
	\begin{center}
		\includegraphics[width=\linewidth]{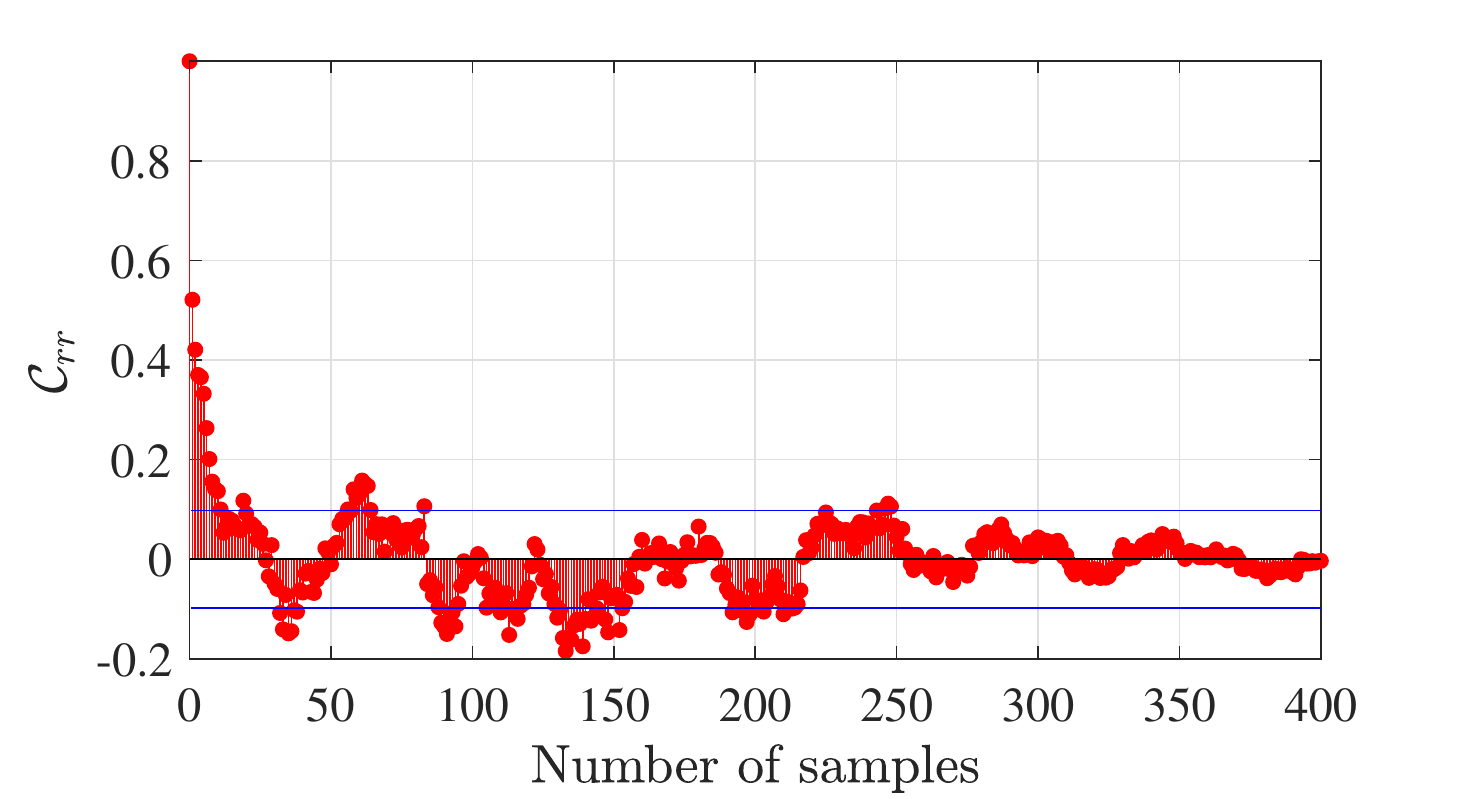}    
	\caption{Plot of auto-correlation of the residual sequence  at location of A2, showing that the samples are correlated.} 
		\label{fig:Autocor_railpad}
	\end{center}
\end{figure}
The $\mathcal{G}$-GLRT detector was designed based on the iid assumption of the detection signal $\mathbf{r}$. To examine the validity of this assumption, the auto-correlation $\mathcal{C}_{rr}$ of the residual sequence $\textbf{r}$ is calculated and shown in Fig.~\ref{fig:Autocor_railpad}. It can be seen that the residual sequence is not white and independently distributed, indicating that the applied detection scheme does not meet the iid assumption. Therefore, calculating a threshold based on \eqref{eq:PFA} leads to a higher number of false alarms and a detector which is sub-optimal. Moreover, the performance of the detector is significantly sensitive to the change in statistic over time. Hence, the threshold $\gamma_g$ providing the desired false alarm rate is computed by using the cumulative density function (CDF) of the test statistics under the hypothesis $\mathcal{H}_0$, 
\begin{equation}
P_{FA}=\int_{\{g:g>{\gamma}_g\}} p(g;\mathcal{H}_0 ) \, \mathrm{d}{g}. \label{eq:PFA_g}
\end{equation}
In the latter equation, $p(g;\mathcal{H}_0)$ is the density function which can well characterize the behavior of $g(k_w)$ at location A2 where the railpad is considered to be in healthy condition.

\begin{figure}[tbp]
	\begin{center}
		\includegraphics[width=\linewidth]{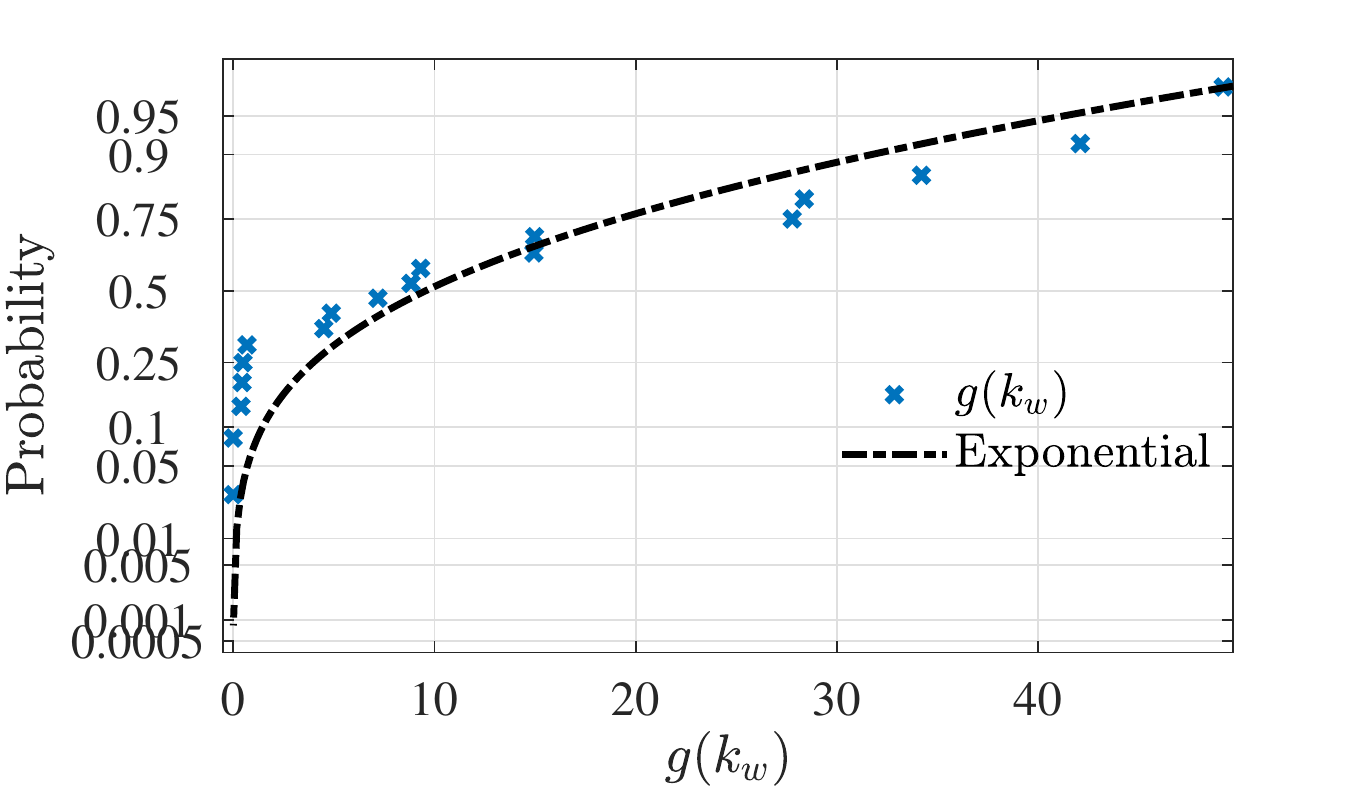}    
	\caption{GLRT test statistics $g(k_w)$ fitted against an exponential distribution.} 
		\label{fig:gkprob}
	\end{center}
\end{figure}

Figure~\ref{fig:gkprob} shows the estimation of $p(g;\mathcal{H}_0)$ from the data $g(k_w)$. It is found that an exponential distribution with the PDF, 
\begin{equation}
p(g;\mathcal{H}_0)=\frac{1}{\alpha_{0g}} \exp\left(-\frac{g}{\alpha_{0g}}\right)  \quad \text{for} \quad \alpha_{0g}> 0,\quad g\geq0  
\label{eq:PgH0}
\end{equation}
is an appropriate fit to the data. Inserting \eqref{eq:PgH0} into \eqref{eq:PFA_g} the following equation can be derived
\begin{equation}
    1-P_{FA}=\left(1-\exp(-\frac{\gamma_g}{\alpha_{0g}})\right)H(\gamma_g),
\end{equation}
where $H(\cdot)$ is the Heaviside step function. Having the estimation of the parameter of the exponential distribution $\alpha_{0g}$, the threshold for the desired probability of false alarm can be calculated as,
\begin{equation}
\gamma_g=-\ln(P_{FA})\alpha_{0g}.
\label{eq:gamma_g}
   \end{equation}
For the monitoring system to trigger a false alarm every 12 months (i.e., $P_{FA}=0.083$), the exponential fit shown in Fig.~\ref{fig:gkprob} gives a threshold $\gamma_g=39.2$. 

\subsection{Implementation}
\begin{figure}[tbp]
	\begin{center}
		\includegraphics[width=\linewidth]{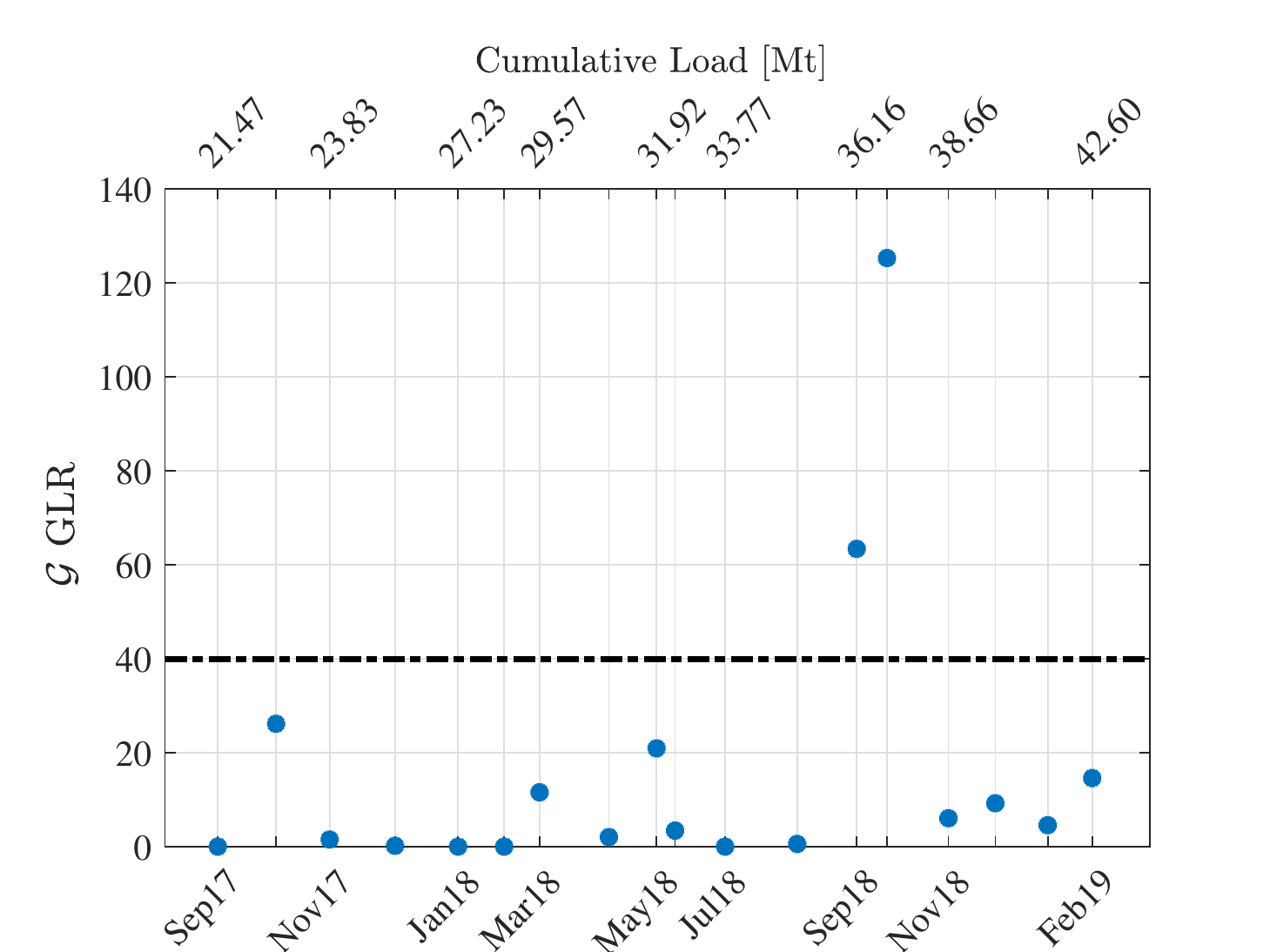}    
	\caption{Performance of the monitoring system for evaluating the railpad status at location A2. Window size one month and no overlap.} 
		\label{fig:glrA2pad}
	\end{center}
\end{figure}

\begin{figure}[tbp]
	\begin{center}
		\includegraphics[width=\linewidth]{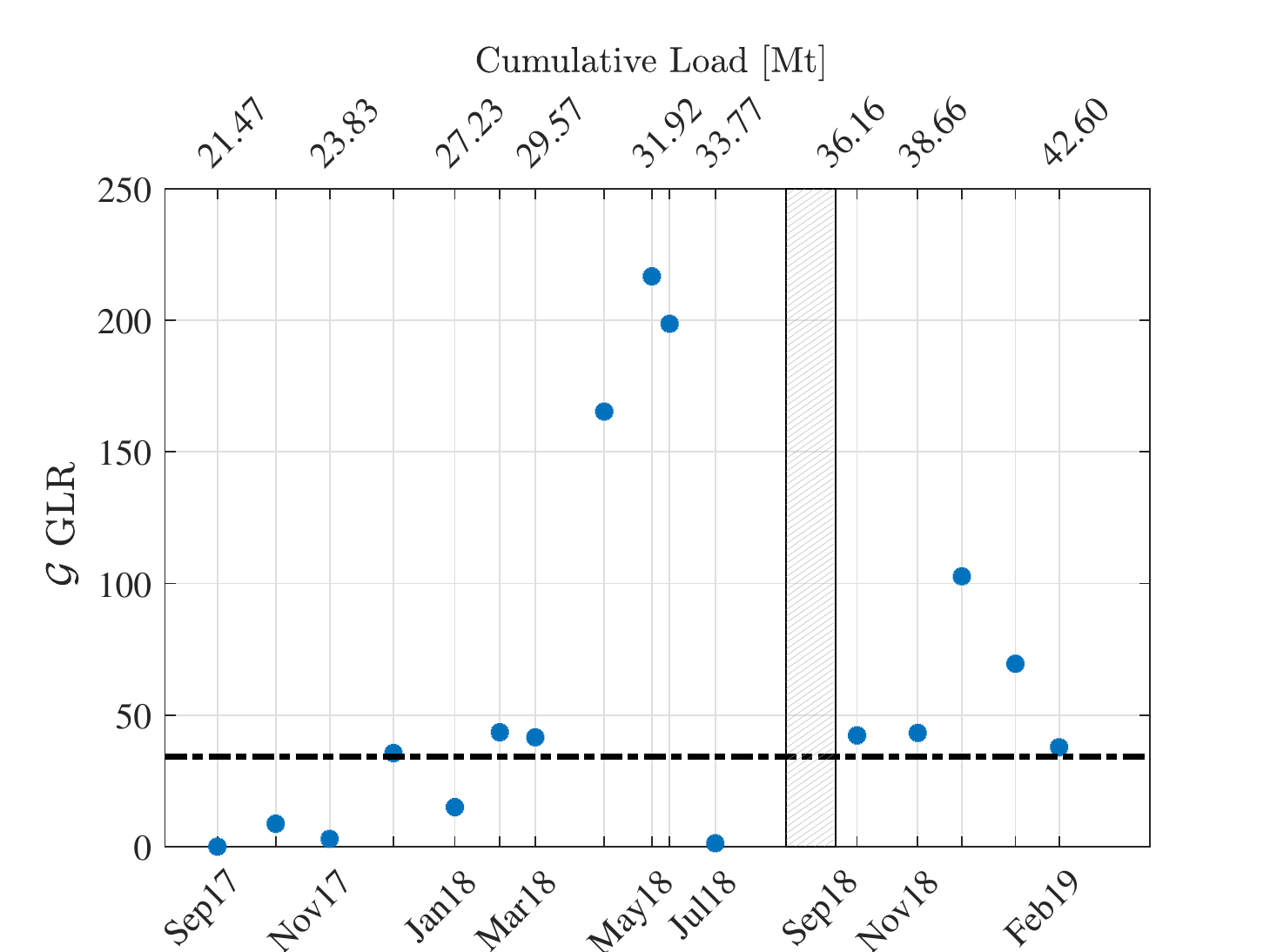}    
	\caption{Performance of the monitoring system for evaluating the railpad status at location A4. Window size one month and no overlap.} 
		\label{fig:glrA4pad}
	\end{center}
\end{figure}

\begin{figure}[tbp]
	\begin{center}
		\includegraphics[width=\linewidth]{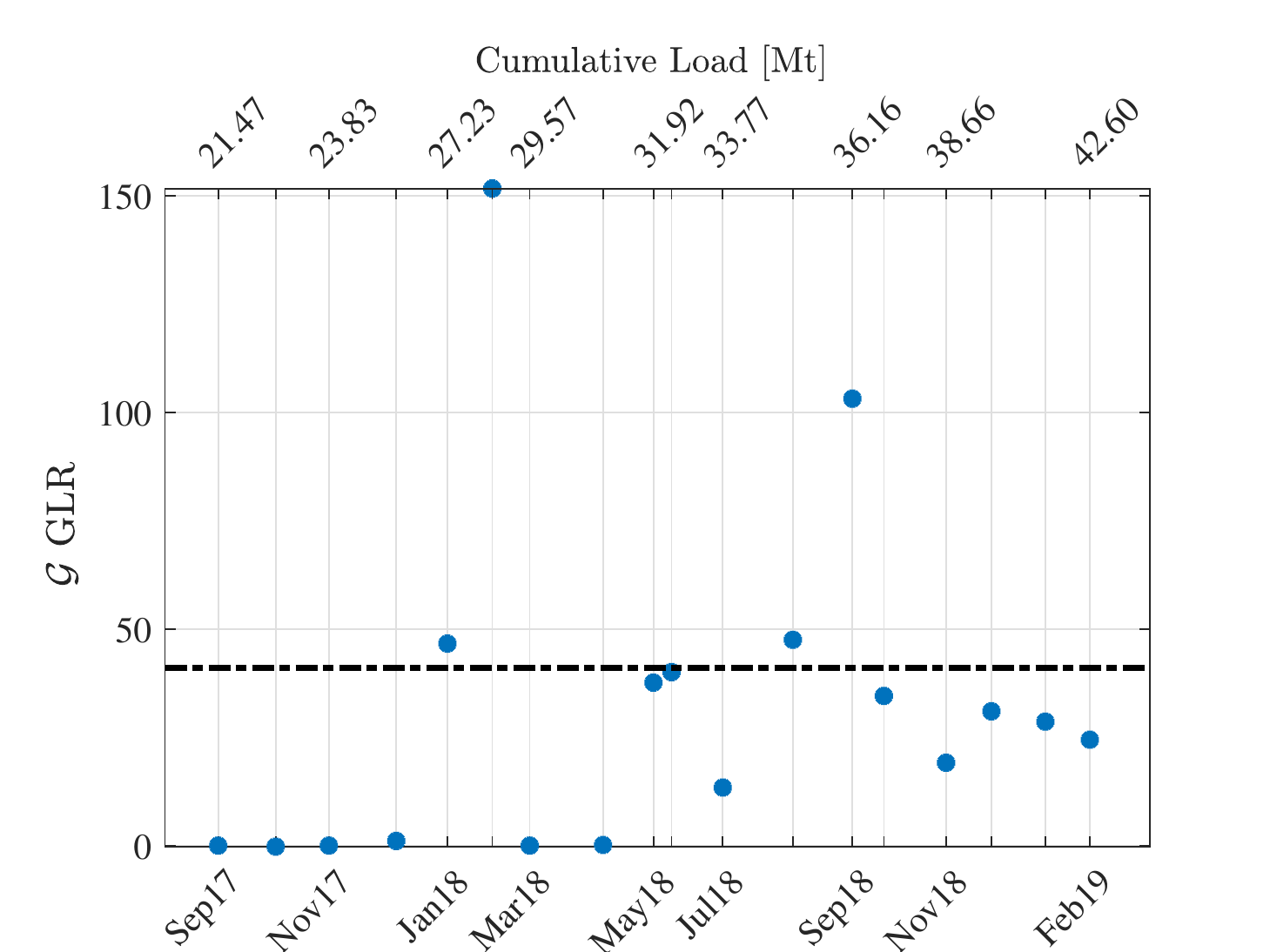}    
	\caption{Performance of the monitoring system for evaluating the railpad status at location A11. Window size one month and no overlap.} 
		\label{fig:glrA11pad}
	\end{center}
\end{figure}

By considering the residual sequence calculated over a period of 18 months, the capability of the designed monitoring system for detecting changes in the status of the railpads at locations A2, A4 and A11 along the turnout is investigated. Figures~\ref{fig:glrA2pad}-\ref{fig:glrA11pad} show the performance of the $\mathcal{G}$-GLRT for the detection and estimation windows equal to one month, and the threshold $\gamma_g=39.2$. The shaded area in Fig.~\ref{fig:glrA4pad} indicates a two-month period in which low-quality data has been collected by the accelerometer mounted at location A4. This problem was solved by re-installing the sensor at this location. Figure~\ref{fig:glrA2pad} shows that the monitoring system has triggered one false alarm for the railpad at location A2 (in August 2018) within the entire period of monitoring. The false alarm is amplified in September 2018 and then withdrawn by the monitoring system in the following month. This observation is in line with the design consideration for the threshold (i.e., one false alarm every 12 months) and, therefore, suggesting no significant variation in the status of the railpad at location A2.

The test statistics for location A11 presented in Fig.~\ref{fig:glrA11pad} shows two false alarms in the considered period, one triggered in January 2018 and one in August 2018. From a probabilistic perspective, occurrence of two false alarms within the period of 18 months can be expected and it is in accordance with the design expectation. The obtained result demonstrates that the alarms have been withdrawn by the monitoring system. Therefore, it can be concluded that the condition of the railpad at location A11 did not changed during the considered period.

It can be seen from the result provided in Fig .\ref{fig:glrA4pad} that the monitoring system has triggered several alarms, indicating a noticeable change in the status of the railpad at location A4. The monitoring system has never withdrawn the alarms after September 2018. Moreover, significant changes are detected from March 2018 to July 2018 which are also visible in the probability plots shown in Fig.~\ref{fig:residualprobA2A4}.

\section{Discussion}\label{sec:discussion}
Currently, there is no inspection tool which can be employed by the track infrastructure managers for evaluating the quality of railpads in service. Moreover, due to the particular configuration of the track infrastructure it is not possible to visually detect defects in the railpad layer placed between rail and sleepers. Therefore, examining the status of in-service railpads is a complicated task which may be accomplished by an appropriate data-driven monitoring tool. The monitoring system here developed provides the possibility of supervising in-service railpads without disassembling the railway track and interrupting railway operations. 

\subsection{Challenges and solutions}
Railpads often exhibit a complex behavior over time since their properties are affected by different factors such as preloading, temperature variations, excitation frequencies and aging. It is essential that the monitoring tool developed for the analysis of railpad quality can distinguish changes in railpad stiffness caused by the aging from those created by other factors. The monitoring tool developed in this paper is capable to minimize the effect of preloading through considering identical trains in the analysis, and the effect of temperature by utilizing a temperature-frequency model obtained for the railpad. The railpad stiffness depends also on the frequency of the applied excitation force. The influence of speed-dependent excitation frequencies (i.e., parametric frequencies) is minimized by considering identical trains with approximately the same speed range. Moreover, a relatively narrow frequency interval in which the second track resonance frequency is expected (i.e. [400 650]Hz \cite{dahlberg2006handbook, barkhordari}) is taken into account in the monitoring system design, which limits the frequency variations.

\subsection{False alarms and threshold}
The value of the threshold for change detection is selected based on a desired rate of probability of false alarms which is considered to be one false alarm every 12 months. Currently, railway infrastructure managers replace worn or defective railpads with new ones based on a periodic maintenance strategy and typically after a certain number of years. Therefore, monthly estimation of the status of railpads provided by the designed monitoring tool is not a must. Moreover, in the case of observing a false alarm triggered by the monitoring system, no immediate action is required. In fact the railway infrastructure manager can instead wait until new data is collected and the new value of the test statistics is provided by the monitoring system. A renewal action can then be executed if the alarm is not withdrawn within an extended period of time (e.g. 3 - 6 months). In the current study, the results obtained for the railpads at locations A2 and A11 (Figs.~\ref{fig:glrA2pad} and \ref{fig:glrA11pad}) indicate that the false alarms are withdrawn by the monitoring system and, therefore, no renewal action is required. However, a different behavior is observed for location A4 in Fig.~\ref{fig:glrA4pad} where the monitoring system does not withdrawn the alarm. This is presumably due to a defect in the railpad layer at this location. However, this hypothesis could not be verified.

\subsection {Portability of the method}
Although the designed monitoring tool is employed to evaluate the health status of railpads along a particular railway turnout, as a data-driven method, it has a great potential to be portable. The method merely requires the track vibration data measured over time. The vibration data can be collected by either a track-side measurement system or by a train equipped with an on-board measurement set-up. For analysis of the data collected by an on-board measurement system, some modifications to the proposed track resonance estimation scheme might be required in order to distinguish between train and track dynamics in the measured data.

\section{Conclusions} \label{sec:conclusion_railpad}
A novel monitoring tool for the long-term performance analysis of in-service railpads was designed and tested in this paper. The monitoring tool is a data-driven system relying on train-induced vibration data collected by a track-side measurement system. The proposed monitoring method consists of three steps: track resonance frequencies estimation, temperature-frequency model derivation, and statistical change detection. To estimate the second track resonance frequency representing the dynamic behavior of railpad, a combination of the EMD technique and the N4SID algorithm was employed. A piecewise linear regression model describing the relation between the ambient temperature and the second track resonance frequency was then obtained. The vector of the estimated frequencies and the obtained temperature-frequency model were used to find a residual sequence in which the temperature effect is minimized. The residual sequence was employed as input to a $\mathcal{G}$-GLRT change detection algorithm to detect changes in the status of the railpad layer over time. It was shown that the proposed monitoring tool is capable of detecting changes in the railpad properties caused by a factor different from temperature and load variations.

\section*{Acknowledgements}
This research was part of the \href{http://www.intelliswitch.dk/}{INTELLISWITCH} project. The research is financially supported by Innovation Fund Denmark under grant number 4109-00003B. The authors gratefully acknowledge this support. The authors also thank the Danish Meteorological Institute for the provision of meteorological data.

\bibliographystyle{Bib/IEEEtran}
\bibliography{Bib/IEEEabrv,Bib/IEEEbibpad}
\end{document}